%% file: main.tex
\pdfoutput=1
\documentclass[sigconf]{acmart}

\copyrightyear{2017}
\acmYear{2017}
\setcopyright{acmcopyright}
\acmConference{CCS '17}{October 30-November 3, 2017}{Dallas, TX,
USA}\acmPrice{15.00}\acmDOI{10.1145/3133956.3133990}
\acmISBN{978-1-4503-4946-8/17/10}

\usepackage{booktabs} %
\usepackage{epsfig,endnotes}
\usepackage{amsmath}
\usepackage{url}
\usepackage{multirow}
\usepackage{color}
\usepackage{balance}
\usepackage{algorithm}
\usepackage[noend]{algpseudocode}
\usepackage{caption}
\usepackage{csquotes}
\usepackage{graphicx}
\usepackage{subfigure}
\usepackage{enumitem}

\definecolor{navy}{RGB}{0,0,128}
\definecolor{forest}{RGB}{69,139,0}

\newcommand{\eg}{{\em e.g.,\ }}
\newcommand{\ie}{{\em i.e.\ }}
\newcommand{\etc}{{\em etc.}}
\newcommand{\etal}{{\it et al.\ }}

\newcommand{\subsubsecspace}{\vspace{0.05in}}

\newcommand{\subsubsectitle}[1]{\subsubsecspace\noindent\textbf{#1}}
\algnewcommand{\LeftComment}[1]{\Statex \(\triangleright\) #1}

\newenvironment{packed_itemize}{
	\begin{list}{\labelitemi}{\leftmargin=2em}
		\setlength{\itemsep}{1pt}
		\setlength{\parskip}{0pt}
		\setlength{\parsep}{0pt}
		\setlength{\headsep}{0pt}
		\setlength{\topskip}{0pt}
		\setlength{\topmargin}{0pt}
		\setlength{\topsep}{0pt}
		\setlength{\partopsep}{0pt}  
	}{\end{list}}

\newenvironment{packed_enumerate}{
	\begin{enumerate}
		\setlength{\itemsep}{3pt}
		\setlength{\parskip}{0pt}
		\setlength{\parsep}{0pt}
		\setlength{\headsep}{0pt}
		\setlength{\topskip}{0pt}
		\setlength{\topmargin}{0pt}
		\setlength{\topsep}{0pt}
		\setlength{\partopsep}{0pt}
	}{\end{enumerate}}

\begin{document}
\title{Automated Crowdturfing Attacks and Defenses in \\Online Review Systems}

\author{Yuanshun Yao}
\email{ysyao@cs.uchicago.edu}
\affiliation{University of Chicago}
\author{Bimal Viswanath}
\email{viswanath@cs.uchicago.edu}
\affiliation{University of Chicago}
\author{Jenna Cryan}
\email{jennacryan@cs.uchicago.edu}
\affiliation{University of Chicago}
\author{Haitao Zheng}
\email{htzheng@cs.uchicago.edu}
\affiliation{University of Chicago}
\author{Ben Y. Zhao}
\email{ravenben@cs.uchicago.edu}
\affiliation{University of Chicago}

\begin{abstract}
  Malicious crowdsourcing forums are gaining traction as sources of spreading
  misinformation online, but are limited by the costs of hiring and managing
  human workers. In this paper, we identify a new class of attacks that
  leverage deep learning language models (Recurrent Neural Networks or RNNs)
  to automate the generation of fake
  online reviews for products and services. Not only are these attacks
  cheap and therefore more scalable, but they can control rate of content output to
  eliminate the signature burstiness that makes crowdsourced campaigns easy to
  detect.

  Using Yelp reviews as an example platform, we show how a two phased review
  generation and customization attack can produce reviews that are
  indistinguishable by state-of-the-art statistical detectors. We conduct a
  survey-based user study to show these reviews not only evade human
  detection, but also score high on ``usefulness'' metrics by users. Finally,
  we develop novel automated defenses against these attacks, by leveraging
  the lossy transformation introduced by the RNN training and
  generation cycle. We consider countermeasures against our mechanisms, show
  that they produce unattractive cost-benefit tradeoffs for attackers, and
  that they can be further curtailed by simple constraints imposed by online
  service providers.
\end{abstract}




\maketitle

\input{intro}
\input{background}

\input{rnn}

\input{details}

\input{generation}

\input{eval}
\input{defense}
\input{related}
\input{discussion}
\appendix
\input{appendix}

\section*{Acknowledgments}
We wish to thank our anonymous reviewers for their constructive
feedback, and William Yang Wang for insightful discussions.
This project was supported by NSF grants CNS-1527939 and
CNS-1705042. Any opinions, findings, and conclusions or recommendations
expressed in this material are those of the authors and do not necessarily
reflect the views of any funding agencies.

\bibliographystyle{ACM-Reference-Format}
\bibliography{refs,zhao}

\end{document}

%% file: intro.tex
\section{Introduction}

The Internet is no longer the source of reliable information it once was. Today,
misinformation is being used as a tool to harm competitors, win political
campaigns, and sway public opinion. Clashes between conflicting accounts
occur daily on social networks and online discussion forums, and the
trustworthiness of many online information sources is now in question.

One highly effective weapon for spreading misinformation is the use of
crowdturfing campaigns~\cite{crowdturfing_2,crowdturfing_3,crowdturfing_4},
where bad actors pay groups of users to perform questionable or illegal
actions online. Crowdturfing marketplaces are the corrupt equivalents of
Amazon Mechanical Turk, and are rapidly growing in China, India and the
US~\cite{crowdturfing_2,crowdturfing_4}.  For example, an attacker can pay
workers small amounts to write negative online reviews for a competing
business, often fabricating nonexistent accounts of bad experiences or
service. Since these are written by real humans, they often go undetected by
automated tools looking for software attackers.

Thankfully, two factors limit the growth and impact of crowdturfing
campaigns. First, they require monetary compensation for each task
performed. Larger campaigns can incur significant cost on an attacker, and
this limits the scale of many campaigns. Second, the predictable reaction of
workers often produce actions (and output) synchronized in time, which can be
effectively used as a feature to classify and identify crowdturfing
campaigns~\cite{crowdturfing_2,crowdturfing_4}. A more knowledgeable attacker
can apply adversarial techniques ({\em e.g.} poisoning training data,
targeted evasion) against machine learning (ML) classifiers, but such
techniques have limited impact, and require significant coordination across
workers~\cite{crowdturfing_5}.

But just as ML classifiers can effectively detect these attacks, advances in
deep learning and deep neural networks (DNNs) can also serve to make these
attacks much more powerful and difficult to defend. Specifically, we believe
that in limited application contexts, DNNs have reached a point where they
can produce sufficiently clear and correct content effectively
indistinguishable from those produced by humans. To illustrate our point, we
focus on the domain of online reviews for e-commerce products and services,
where millions of users upload reviews to sites such as Yelp, TripAdvisor and
Amazon. Online reviews tend to be short, and focused on a limited range of
topics defined by the application domain, {\em e.g.} quality of food and
service at a restaurant. We believe that well designed and tuned DNNs are now
capable of producing realistic online reviews. If successful, attack
campaigns using DNN-based fake reviews would be much more powerful, because
they are highly scalable (no per-review payments to human writers) and harder
to detect, as scripts can control the rate of review generation to avoid the
telltale burstiness that makes crowdsourced reviews so easily
detectable~\cite{crowdturfing_2,crowdturfing_5}.  

In this paper, we identify a class of attacks based on DNN-based fake review
generation. We demonstrate that DNN-based review generators are practical and
effective, using a combination of ML-trained review generation and
context-based customization. Using Yelp restaurant reviews as a target
platform, we show empirically that synthetic reviews generated by our tools
are effectively indistinguishable from real reviews by state-of-the-art
detectors relying on linguistic features. We carry out a user study ($N$=600)
and show that not only can these fake reviews consistently avoid detection by
real users, but they provide the same level of user-perceived ``usefulness''
as real reviews written by humans.

We then examine potential defenses, and propose an ML-classifier based defense
that leverages the inherent computational limitations of most RNNs against
the attacker. This leverages the fact that generative language models build
fixed memory presentations of the entire training corpus, which limits the
amount of information that can be captured from the training corpus.  We show
that the cycle of processing real reviews through a RNN-based model training
and text generation is lossy, and the resulting loss can be detected by
comparing the character level distribution of RNN-generated reviews against
those written by real users. We also consider potential countermeasures, and
show that increasing model complexity produces diminishing returns in
evasion, while resource costs increase dramatically. 

In summary, our work produces several key takeaways:
\begin{packed_enumerate}
\item We demonstrate the feasibility of automated generation of product
  reviews for online review sites, using a RNN-based approach for review
  generation and customization. Our key insight is that while automated
  generation of arbitrary length content is challenging, generation of
  shorter text in fixed application domains is practical today.
\item We show that RNN-based synthetic reviews are robust against state of
  the art statistical and ML-based detectors.  In addition, our user-study
  shows they are largely indistinguishable from real reviews to human
  readers, and appear to provide similar levels of ``usefulness'' utility as
  determined by readers.
\item We propose a novel defense that leverages the information loss inherent
  in an RNN training process to identify statistically detectable variations
  in the character-level distribution of machine-generated reviews.  We show
  that our defense is robust against countermeasures, and that avoiding
  detection involves the attacker paying rapidly accelerating costs for
  diminishing returns. 
\end{packed_enumerate}

We believe this is a practical new attack that can have significant impact on
not only user-generated review sites like Yelp, but potential attacks on
content generation platforms such as Twitter and online discussion forums. We
hope these results will bring attention to the problem and encourage further
analysis and development of new defenses.

%% file: background.tex
\section{Preliminaries}
\label{sec:background}

We begin our discussion with background material on online review systems,
and content generation based on deep learning networks (RNNs in
particular). For simplicity, we focus our discussion on online review systems
such as Yelp, Amazon and TripAdvisor.

\subsection{Crowdsourced Attacks on Review Systems}
Most popular e-commerce sites today
rely on user feedback to rate products, services, and online content.
Crowdsourced feedback typically includes a \textit{review} or opinion,
describing a user's experience of a product or
service, along with a rating score (usually on a 1 to 5 scale).

Unfortunately, many review systems are plagued by \textit{fake
reviews}, \eg Yelp~\cite{yelp_fake_business},
Amazon~\cite{fake_review_media_1}, iTunes~\cite{fake_review_media_2} and
TripAdvisor~\cite{tripadvisor_fake_news},
where an attacker manipulates
crowd opinion using fake or deceptive reviews. To boost their reputation or
to damage that of a competitor, businesses can solicit fake reviews that
express an overly positive or negative opinion about a 
business~\cite{crowdturfing_2, crowdturfing_4,yelp_fake_business}. 
Studies on Yelp found that a one star rating increase for restaurants can lead to a 
5--9\% boost in revenue~\cite{luca2016fake}.

Sites like Yelp and Amazon have been consistently engaged in a cat and mouse
battle with fake reviews,  as attackers try to adapt and bypass
various defense schemes~\cite{fake_review_media_4}. 
Yelp's \textit{review filter} system flags suspicious reviews
and even raises an alert to the consumer if a business is suspected of
engaging in large-scale opinion manipulation~\cite{yelp_filter}.

Recently, attacks have been known to generate highly deceptive (authentic
looking) fake reviews written by paid users.  Much of this comes from
malicious crowdsourcing marketplaces, known as \textit{crowdturfing} systems,
where a large pool of human workers provide on-demand effort for
completing various malicious tasks~\cite{yelp_crowdsouring_site_1,
  yelp_crowdsouring_site_2}.  In the next section, we introduce an attack
powered by an AI program that can replace human writers and achieve high
attack success.

\subsection{Our Attack Model}

We assume the attacker's goal is to use an AI program to generate fake reviews 
that are indistinguishable from real reviews written by human users. We only
focus on the generation of review text, which is crucial to deceive users and
to manipulate their opinion. We do not consider the manipulation of 
metadata associated with a review or reviewer. Metadata can include any
information other than the textual content, \eg reviewer reputation, review
history, posting date and IP address.

\subsubsecspace
\noindent \textbf{Key Insight.} There have been significant recent advances
in building probabilistic \textit{generative language models} on Neural
Networks, specifically \textit{Recurrent Neural Networks (RNNs)}.  Even
trained on large datasets, RNNs have generally fallen short in producing
writing samples that truly mimic human
writing~\cite{longrnndifficult}. However, our insight is that 
the quality of RNN-generated text is likely more than sufficient for
applications relying on domain-specific, short length user-generated content,
\eg online reviews.

\subsubsecspace
\noindent \textbf{Assumptions.} 
\begin{packed_itemize}
\item The attacker has access to a corpus of real reviews to train the
generative language model. Popular sites like Yelp have already released large
review datasets~\cite{yelp_official}. Attackers can also download reviews, or
large review datasets made public by researchers~\cite{amazon_data,
  movieLens, ibdm_dataset}.

\item The attacker has knowledge of the domain of a product (\eg cameras) or
business (\eg restaurants, clothing stores) which allows training on a review 
corpus that matches the domain.

\item The attacker has access to sufficient computational resources 
for training neural networks. Today, commodity GPU machines can efficiently train 
DNNs, and can be purchased or accessed in the Cloud~\cite{amazon_gpu}. More
recently, building and training an ML model in the Cloud has become easier
with the emergence of Machine Learning as a Service platforms~\cite{mlaas_imc}.
\end{packed_itemize}

\begin{figure}[!t]
	\centering
	\mbox{
		\subfigure[Human-based attack.]
		{\includegraphics[width=0.22\textwidth]{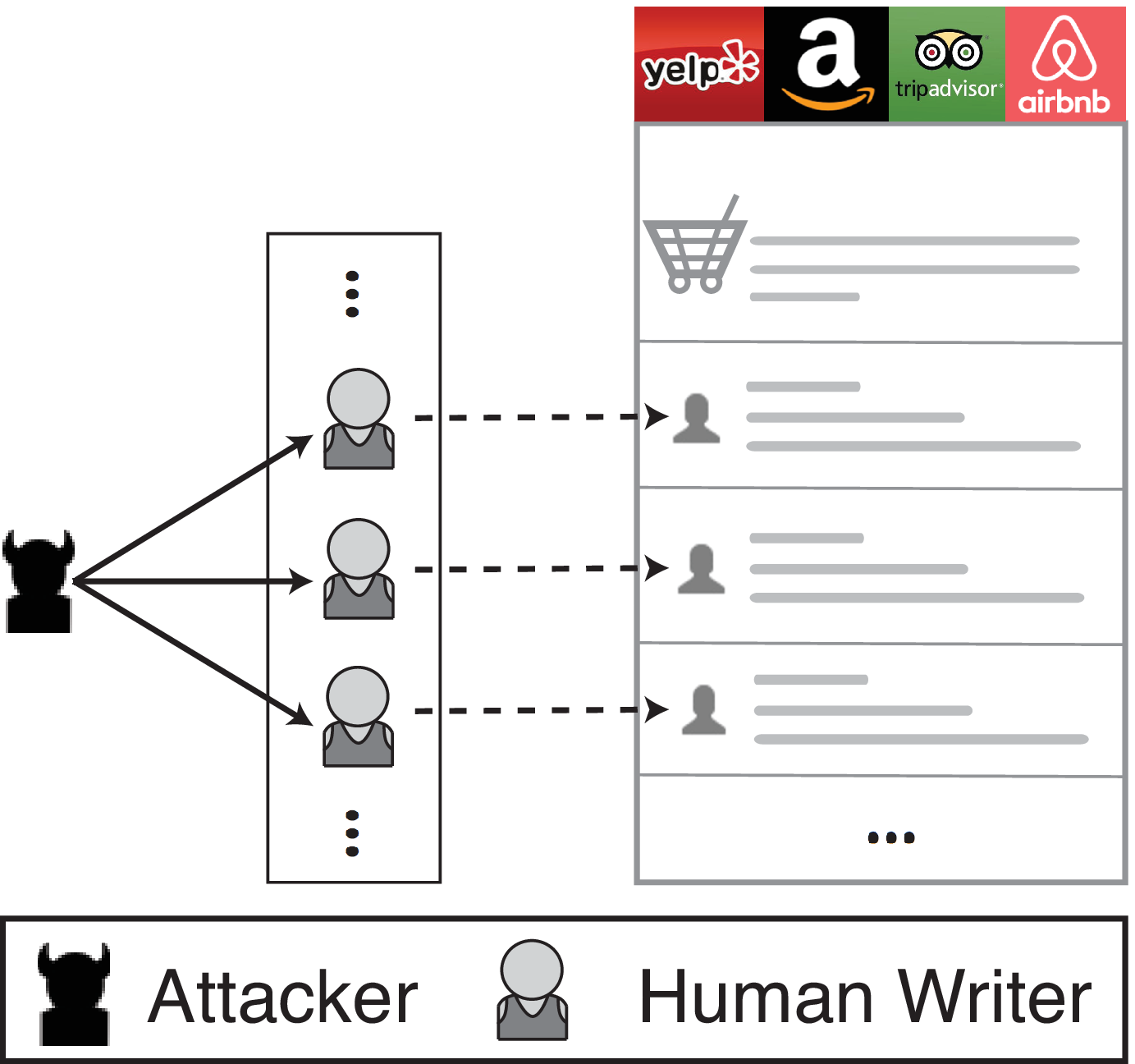}}
		\hspace{0.01in}
		\subfigure[Machine-based attack.]
		{\includegraphics[width=0.22\textwidth]{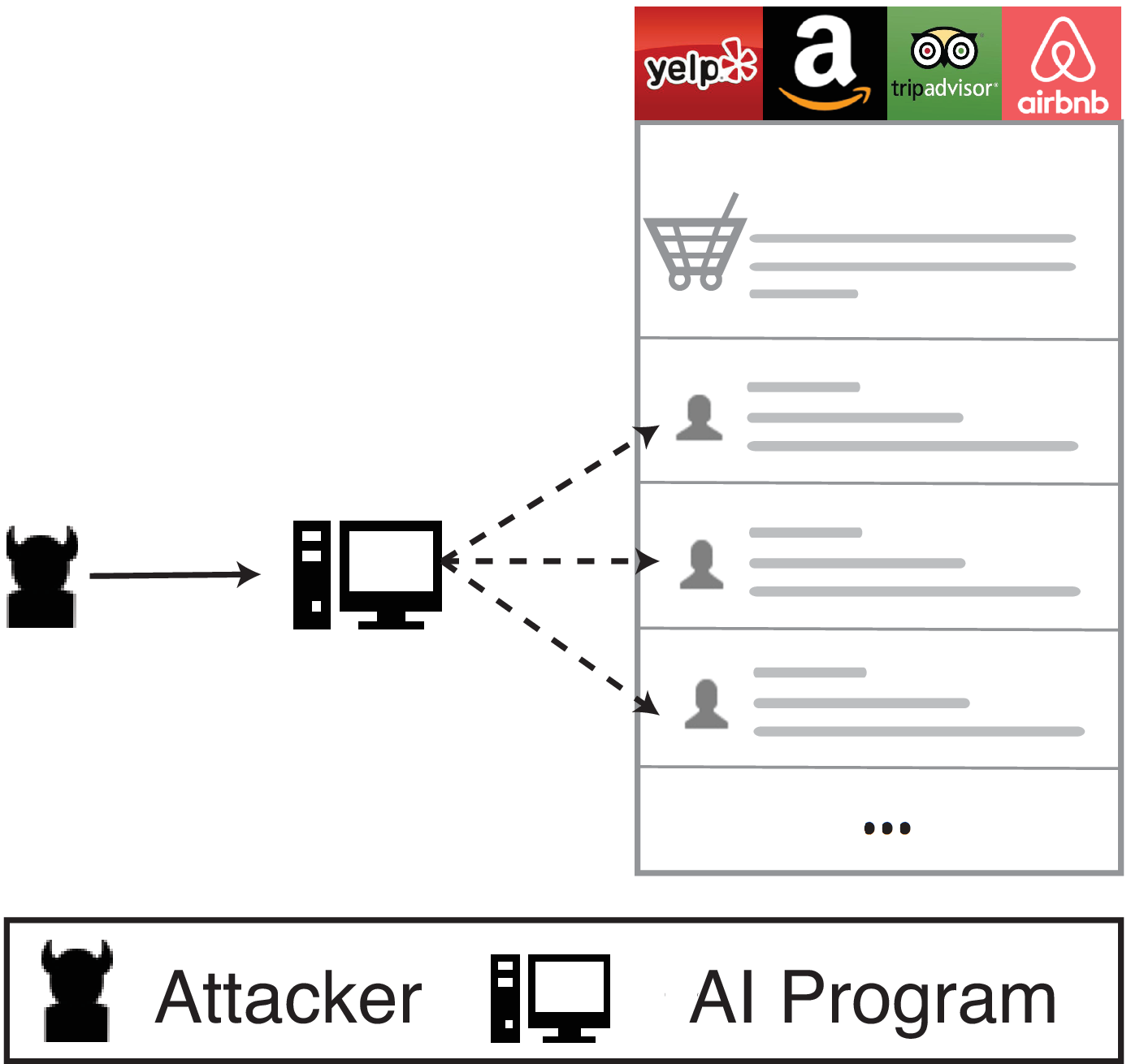}}       }
	\caption{Fake review attack: Human-based vs. Machine-based.}
	\label{fig:overview}
\end{figure}

%% file: rnn.tex
\subsection{RNNs vs. Crowdsourced Authors}

Traditional attacks using fake reviews typically hire human writers to write
reviews. Instead, our work considers automated, machine-based review attacks
leveraging DNN-based language models (see Figure~\ref{fig:overview}). 
Here, we compare the two approaches and highlight the benefits of automated
review attacks.

The key difference between these two approaches is the quality of writing in
the generated text content. To influence user opinion and alter decisions,
fake reviews need to be written in such a way to mimic content written by
real users. Broken grammar, misspellings and broken context can make a review
appear fake. Existing machine-based text generations techniques, such as
\textit{n-gram models} and \textit{template-based models} are known to have
limitations in generating realistic-looking text~\cite{rnn_thesis,
  rule_based_sucks}.  Hence, the reviews generated based on those techniques
are likely to be identified as fake by readers~\cite{dspin}.  In contrast, a
generative RNN model can generate much more coherent
text~\cite{arisoy2012deep, mikolov2011empirical}, but still falls short for
larger types of content~\cite{longrnndifficult}. 

There are some obvious benefits to using a software-controlled RNN to
generate fake reviews. First, it removes the cost of paying human
writers, which costs \$1-\$10 per review on Yelp according to prior 
work~\cite{yelpavgcost}. To further obtain an independent estimate, 
we search Blackhat SEO Forum\footnote{\url{https://www.blackhatworld.com/}} for
``Yelp reviews,'' and observe an average price of \$19.6 per review based on a
random sample of 20 posts.
Perhaps more
importantly, software generators can control the specific timing and output
of reviews, so they are harder to detect.  When attackers launch large-scale
fake review campaigns using human writers, workers tend to rapidly generate the 
requested reviews,
producing a burst in new reviews that is easily
detected~\cite{crowdturfing_2,crowdturfing_5}.

\subsection{RNN as a Text Generative Model}
\label{sec:rnn_generative}

We provide background on text generation using Recurrent Neural Networks.

\begin{figure}[!t]
\centering
\includegraphics[width=0.465\textwidth]{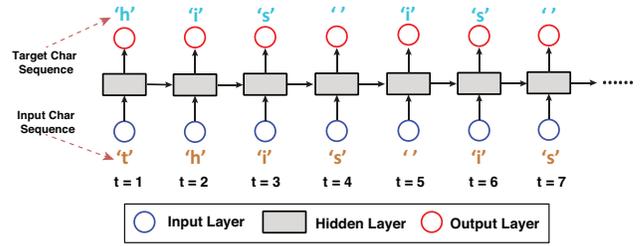}
\caption{RNN generative model training.}
\label{fig:gen_rnn}
\end{figure}

Neural Networks are computational models that use a connected network of
\textit{neurons} to solve machine learning tasks. Neurons serve as the basic
computational units in a Neural Network. In this work, we focus on a specific
class of Neural Networks known as Recurrent Neural Networks (RNNs), which are
better suited for sequential data. RNNs can learn from a large corpus of 
natural language text (character or word sequences), to
generate text at different levels of
granularity, \ie at the character level or word level. We focus on a
character-level RNN due to its recent success on generating high quality 
text~\cite{graves2013generating, gen_rnn_4}. Also, the memory and computational 
cost 
required to train a character-level RNN is lower than word-level RNN since 
number of words is 
significantly larger than number of valid characters in the English 
Language. 

\begin{figure*}[!h]
	\begin{center}
		\includegraphics[width=0.8\textwidth]{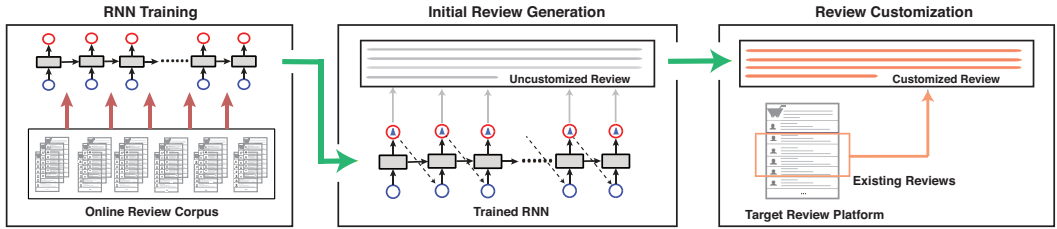}
		\caption{Overview of our attack methodology.}
		\label{fig:sys_design}
	\end{center}
\end{figure*}

\subsubsectitle{RNN Training.} As mentioned before, traditional language
models (\eg n-gram models) exhibit limited performance when trained on long
text sequences since they are able to look back only a few steps of the
sequence. RNN solves this problem by building a more sophisticated ``memory''
model which maintains long term information about what it has seen so far. In
an RNN, ``memory'' is a set of high dimensional weights (hidden states)
learned during the training stage to capture information about all characters
seen in the training sequence. 

Figure~\ref{fig:gen_rnn} illustrates the training process of an
RNN-based text generation model. At each time step $t$, a new character
$\text{x}_\text{t}$ is fed as input to a memory unit of the RNN that maintains a
hidden state $\text{h}_\text{t}$, and provides an output $\text{o}_\text{t}$.
Then it compares the current output $\text{o}_\text{t}$ with the desired output,
which is the next character in the text. The error between them is computed and
the hidden states are updated towards the direction where the error is
minimized. After multiple iterations of updates, the hidden layer will
eventually capture the relationship between each input character and all
characters prior to it, \ie the conditional probability distribution
$P(\text{x}_\text{t+1}$|($\text{x}_\text{1},\dots,\text{x}_\text{t}))$.

\subsubsectitle{Text Sampling.} After an RNN model is trained, text can be
generated by feeding a character, say $\tilde{\text{x}_{\text{0}}}$, to the 
trained RNN. The
RNN returns %
a probability distribution that defines which characters are
likely to occur next, \ie 
$P(\tilde{\text{x}_\text{1}}|\tilde{\text{x}_\text{0}})$. We %
stochastically sample from this distribution to obtain the next character
$\tilde{\text{x}_\text{1}}$. Next, by feeding $~\tilde{\text{x}_\text{1}}$ back 
in to the 
RNN, we %
obtain another probability distribution predicting the next character, \ie
$P(\tilde{\text{x}_\text{2}}|\tilde{\text{x}_\text{0}}, 
\tilde{\text{x}_\text{1}})$. This 
process can 
be repeated
to generate continuous text $\tilde{\text{x}_\text{0}}, 
\tilde{\text{x}_\text{1}}, 
\tilde{\text{x}_\text{2}},\ldots, \tilde{\text{x}_\text{N}}$.

\subsubsectitle{Temperature Control. } An important parameter that we can
manipulate during the sampling stage is \textit{temperature}. %
Temperature is a parameter used in the
\textit{softmax} function during the sampling stage when converting the output
vector $\text{o}_\text{t}$ to a probability distribution. Formally:
\begin{equation}
P(\text{x}_\text{t}|(\text{x}_\text{1},\ldots,\text{x}_\text{t-1})) = 
\text{softmax(o}_\text{t})
\end{equation}
each $o_{t}$ is a N-dimentional vector where N is the size of the character
vocabulary. The \textit{softmax} function is defined as:
\begin{equation}
\label{eqn:softmax} P(\text{softmax}(\text{o}_\text{t})=\text{k}) =
\frac{\text{e}^{{\text{o}_\text{t}^{\text{k}}}/{\text{T}}}}{\sum_{\text{j=1}}^{\text{N}}
 \text{e}^{{\text{o}_{\text{t}}^{\text{j}}}/{\text{T}}}}
\end{equation}
Here, $\text{o}_{\text{t}}^{\text{k}}$ represents the component of the output
corresponding to the character class $k$ (in the vocabulary), at time
$t$, for a given temperature, $T$. 

Temperature controls the ``novelty'' of generated text. Temperatures lower
than 1 amplifies the difference in the sampling probability for each
character. In other words, this reduces the likelihood of the RNN to pick
characters with lower probabilities, in preference of more 
common characters. As a result, this constrains the sampling, and generates
less diverse text, and more potentially repetitive patterns.
As temperature increases, the variation of
sampling probability for each character diminishes, and the RNN will generate
more ``novel'' and diverse text. But along with diversity comes a higher risk
of mistakes (\eg spelling errors, context inconsistency errors \etc).

%% file: details.tex
\section{Attack Methodology and Setup}
\label{sec:core_ideas}

We focus our study on Yelp, the most popular site for collecting and sharing
crowdsourcing user reviews. Yelp's review system is representative of other
review systems, \eg Amazon or TripAdvisor.  In this section, we describe 
details of our
attack methodology, datasets and training setup.

\subsection{Attack Methodology}
\label{subsec:method}

Our attack methodology is illustrated in Figure~\ref{fig:sys_design}. At a high
level, the attack consists of two main stages: (1) The first stage starts by
training a generative language model on a review corpus. The language model is
then used to generate a set of initial reviews. (2) In the second stage, a
customization component further modifies these reviews to capture specific
information about the target entity (\eg names of dishes in a seafood
restaurant), and produces the final targeted fake review. In our experiments,
the customizable content is extracted from a {\em reference dataset}, composed 
of existing reviews associated with the target entity. If there are no existing 
reviews, an attacker can build a reference dataset using reviews of entities in the 
same category (\eg seafood restaurants) as the target. Restaurant category metadata 
is available on Yelp and similar sites, and can be used to identify similar 
entities.

\subsubsectitle{Generating Initial Reviews.} First, the attacker chooses a training
dataset that matches the domain of the target entity. For example, to generate 
reviews targeting restaurants, the attacker would choose a dataset of
restaurant reviews. Next, the attacker trains a generative RNN model using the
dataset. Afterwards, the attacker generates review text using the
sampling procedure in Section~\ref{sec:rnn_generative}. Note that the attacker 
is able to generate reviews at different \textit{temperatures}.

\subsubsectitle{Review Customization.} In general, there is no
control over the topic or context (\eg name of a food in a restaurant) generated
from the RNN model, since the text is stochastically sampled based on the
character distribution. To better target an entity (\eg restaurant), we
further capture the context by customizing the generated reviews with
domain-specific keywords.
This is analogous to crowdsourced fake review markets, where workers are
typically provided additional information about the target entity for a writing
task~\cite{yelp_crowdsouring_site_1}. The information consists of specific nouns
(\eg names of dishes) to be included in the written review. Based on this
observation, we propose an automated noun-level word replacement strategy.

Our method works by replacing specific words (nouns) in the initial review
with new words that better capture the context of the target entity. This
involves three main steps:
\begin{enumerate}[leftmargin=*]
\item \textit{Choose the type of contextual information to be captured.} The
attacker first chooses a keyword $C$ that helps to identify the context. For
example, if the attacker is targeting a restaurant, the keyword can be ``food,''
which will capture the food-related context. If the target is an online
electronic accessories store, then the keyword can be ``accessory'' or
``electronics.''
\item \textit{Identify words in reviews of the reference dataset that capture
context.} Next, our method identifies all the nouns in the reference dataset that are relevant to the keyword
$C$. Relevancy is estimated by calculating lexical similarity using
WordNet~\cite{wordnet}, a widely used lexical database that groups English words
into sets of synonyms and measures their concept relatedness~\cite{wordnet_sim}.
We identify a set of words $p$ in the reference
dataset that have high lexical similarity with the keyword $C$, using a
similarity threshold $\text{MIN}_\text{sim}$. The set of words $p$ captures the
context of the target entity.
\item \textit{Identify words in initial reviews for replacement.}
Finally, we find all the nouns in the review set $R$ that are also relevant to
$C$ using the same method in Step 2. We replace them by stochastically sampling
words in $p$ based on the lexical similarity score.
\end{enumerate}

A detailed version of the above algorithm is available in
Appendix~\ref{appendix:context}. Figure~\ref{fig:context_demp} shows an
example of customizing an initial review that has language more suitable for
a Japanese restaurant, to a review more suitable for an Italian
restaurant. The nouns to be replaced in the initial review are
marked in green, and replacement nouns are marked in blue. Note that we
choose this noun-level replacement strategy because of its simplicity, and
there is scope for further improvement of this technique.

\begin{figure}[!t]
\centering
\includegraphics[width=0.445\textwidth]{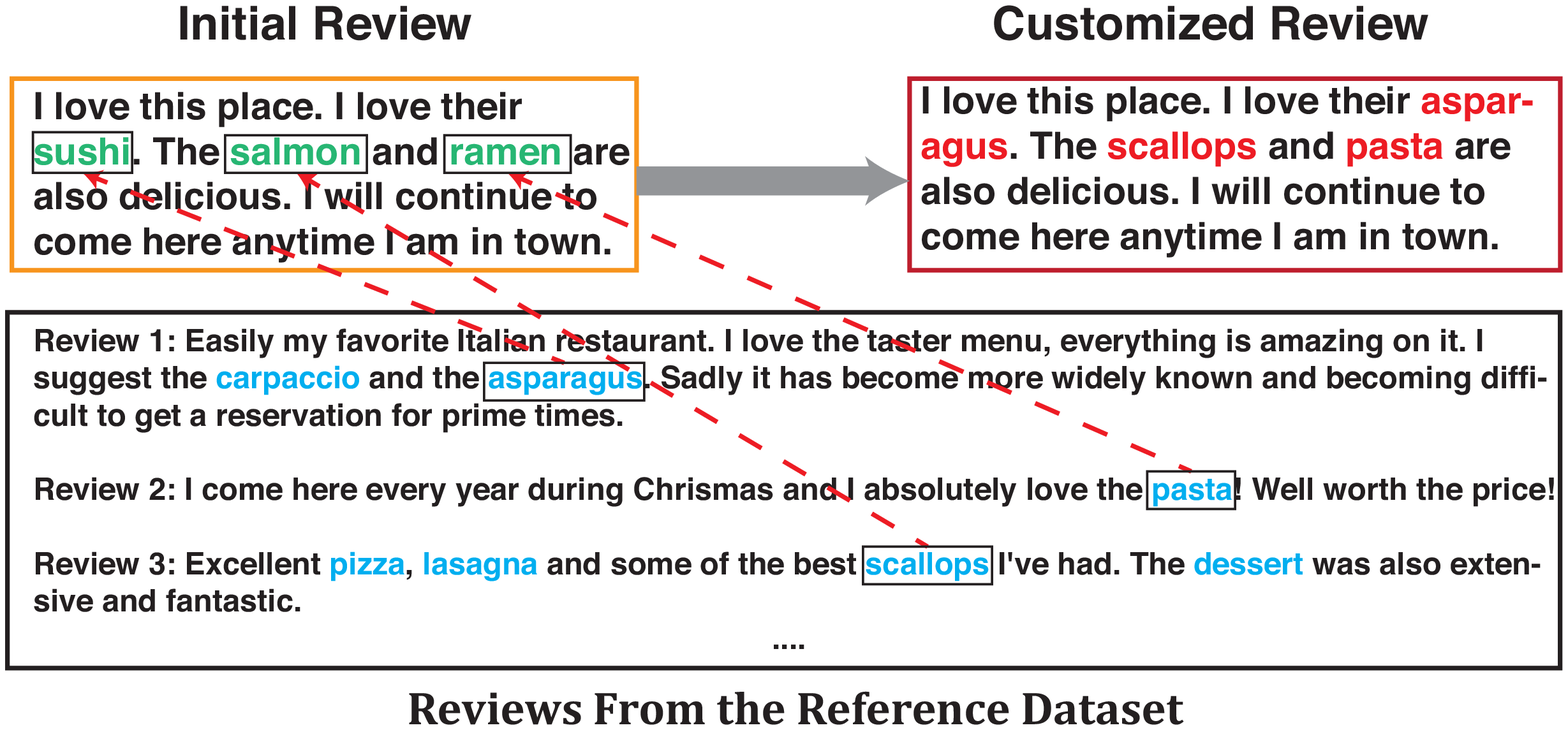}
\caption{Example of review customization.}
\label{fig:context_demp}
\end{figure}

%% file: generation.tex
\subsection{RNN Training and Text Generation}
\label{sec:rnn_modelconfig}
\subsubsectitle{Training Process.} For all experiments, we use a Long
Short-Term Memory (LSTM) model~\cite{lstm}, an RNN variant that has shown
better performance in
practice~\cite{lstm_3}.  %
We examine multiple RNN training configurations used in prior
work~\cite{lipton2015capturing, graves2013generating} and determine the best
configuration empirically through experiments.  The neural network we used
contains 2 hidden layers, each with 1,024 hidden units. For training, the
input string is split into batches of size 256.  Training loss is computed
using \textit{cross-entropy}~\cite{goodfellow2016deep}, and weights are updated using
\textit{Adam}~\cite{adam} optimization, a common optimization technique for
neural network training. The model is trained for 20 epochs
and the learning rate is set to be $\text{2}\times \text{10}^\text{-3}$ and
decays to half of the current rate every time when the loss increases for 5
successive batches. We also monitor training loss and inspect generated
reviews to avoid underfitting or overfitting.

We pre-processed the review text by removing all extra white space and non-ASCII
characters. Additionally, we separate the reviews in the corpus by the 
delimiter tokens ``\textless SOR\textgreater'' (start of review) and ``\textless
EOR\textgreater'' (end of review), so the model also learns when to start and
end a review by generating these two tokens. The RNN is trained using a machine
with Intel Core i7 5930K CPU and a Nvidia TITAN X GPU. The training takes 
$\sim$72 hours. 

\subsubsectitle{Text Generation.} Once we have trained the language model, we
can sample the review text at different temperatures. We generate reviews at 10
different temperatures between 0 and 1: [0.1, 0.2, ..., 1]\footnote{We do not
experiment with temperatures beyond 1.0, because the sampling distribution would
significantly diverge from the true distribution learned from the training
corpus, and lead to overly diverse and incoherent text.}. To start the
text generation process, the model is seeded with the start of review
delimiter token. Conversely, the model identifies the end of a review by
generating the closing delimiter token.

For review customization, we chose the target keyword $C$ to be ``food,'' as
a large number of nouns unsurprisingly relate to food. We set the other
parameter $\text{MIN}_\text{sim}$ to 0.2. After customization, overall, 98.4\%
of the reviews have at least one word replaced. The reviews not affected by
the customization lack suitable content (or words) that capture the context,
including reviews that rarely mention any food or dish in the text, \eg ``I
love this place!  Will be back again!!''

\subsubsectitle{Generated Text Samples.} Table~\ref{tab:review_example} shows
several examples of reviews generated at different temperatures. At higher
temperatures, the RNN is more likely to generate novel content, while at lower
temperatures, the RNN produces repetitive patterns. We include more examples of
generated reviews in Appendix~\ref{appendix:mg_example}.

\subsection{Datasets}
\label{subsec:data}

For our evaluation, we use different datasets of restaurant reviews on Yelp.
Each review in a dataset contains the text of the review, the identity of the
target restaurant and a desired rating score, ranging from one to five
stars. The rating score determines the sentiment of the review and the desired 
textual content.

In the rest of the paper, we present results based on the generation of reviews
tailored towards a five-star rating. This considers the common scenario of an
attack trying to improve the reputation of a restaurant. Our attack methodology
is general and would be the same for other ratings as well.
Appendix~\ref{appendix:mg_example} presents examples of generated reviews
tailored towards one-star and three-star ratings.

Three disjoint datasets of Yelp reviews are used for generating and evaluating 
the attack: a \textit{training},
\textit{ground-truth}, and \textit{attack} dataset. 

\begin{table}
\centering
\resizebox{0.98\columnwidth}{!}{
        \begin{tabular}{|c|c|c|c|}
        \hline
        \textbf{Dataset}  &   \textbf{\begin{tabular}[c]{@{}c@{}}\# of \\
        Restaurants \end{tabular}}   & \textbf{\begin{tabular}[c]{@{}c@{}}\# of
        Fake \\ Reviews (\%)\end{tabular}} &
        \textbf{\begin{tabular}[c]{@{}c@{}}\# of Real \\
        Reviews (\%)\end{tabular}}  \\ \hline
        \textbf{YelpBos}~\cite{yelp_data_1} & 1,028  & 28,151 (22.12\%) & 99,117
        \\ \hline
        \textbf{YelpSF}~\cite{yelp_data_1} &  3,466  & 90,777  (9.94\%) &
        822,772                               \\ \hline
        \textbf{YelpZip}~\cite{yelp_data_2} & 4,204 & 84,484   (13.76\%) &
        529,569                        \\ \hline
        \textbf{YelpNYC}~\cite{yelp_data_2} & 914  & 37,799   (10.48\%) &
        322,858                         \\ \hline
        \textbf{YelpChi}~\cite{mukherjee2013yelp} & 98  & 8,401 (12.83\%) &
        57,061                          \\ \hline
        \textbf{Total} & 9,710 & 249,612 (11.99\%) & 1,831,377
        \\ \hline
 \end{tabular}
}
 \vspace{0.2in}
 \captionof{table}{Summary of ground-truth dataset.}
 \label{tab:eval_datasets}
\end{table}

\subsubsectitle{Training Datasets.} We use the Yelp Challenge dataset to
train the RNN language model, containing a total of 4.1M reviews by 1M
reviewers, collectively targeting 144K businesses~\cite{yelp_official}. The
dataset covers restaurants in 11 cities, spread across 4 countries. We
extract reviews corresponding to different ratings, and found 
617K reviews with a five-star rating from 27K restaurants. In total, these
five-star reviews contain 57M words and 304M characters, 
a sufficiently large dataset for training an RNN.

\subsubsectitle{Ground-truth Dataset.} This dataset, listed in Table~\ref{tab:eval_datasets}, comprises of multiple
Yelp review datasets released by researchers.
By providing ground-truth information
about existing fake and real reviews on Yelp, this dataset enables us to
build machine learning fake review classifiers to evaluate our attack success
(Section~\ref{subsec:mach_test}). Similar to previous work~\cite{yelp_data_1,
  yelp_data_2, mukherjee2013yelp}, we treat Yelp \textit{filtered} and
\textit{unfiltered} reviews as ground-truth information for fake and real
reviews. Yelp attempts to filter reviews that are ``fake,
shill or malicious''~\cite{yelp_filter_2}, but acknowledges imperfections in
the accuracy of the filter~\cite{yelp_filter}. However, given this is the
best information currently available and used by many prior studies on fake
reviews, we use it to establish ground-truth. In the rest of the paper, we use 
\textit{fake} and \textit{real} reviews to refer to Yelp 
filtered and Yelp unfiltered reviews, respectively.

For each dataset, we only consider reviews targeting restaurants\footnote{Since
Yelp also includes reviews for non-restaurant business, \eg hair salon and car
service.}. The resulting dataset contains restaurants in NYC, Chicago, SF,
Boston, and several cities in NJ, VT, CT, and PA.

\subsubsectitle{Attack Dataset.} This dataset contains the reviews generated by
our RNN language model. We use the attack dataset to evaluate attack performance
(Section~\ref{sec:eval}) and defense schemes (Section~\ref{sec:defense}).

The datasets contain similar data to the ground-truth dataset, except for
replacing all fake reviews with our machine-generated reviews. Using our
RNN model, we generate reviews targeting each restaurant in the
ground-truth dataset using different temperatures. For
each temperature, we generate as many reviews as fake reviews
from Yelp for each restaurant, \ie 249,612 machine-generated reviews targeting 
9,710 restaurants.

\begin{table*}[t]
\centering
\begin{tabular}{|c|l|}
	\hline
	\textbf{Temperature} & \multicolumn{1}{c|}{\textbf{Generated Review 
	Text}} \\
	 \hline
	0.1                  & \begin{tabular}[c]{@{}l@{}}I love this place! I have 
	been here a few times and have never been disappointed. The service\\  is 
	always great and the food is always great. The staff is always friendly and 
	the food is always\\ great. I will definitely be back and try some of their 
	other food and service.\end{tabular} \\ \hline
	0.5                  & \begin{tabular}[c]{@{}l@{}}I love this place. I have 
	been going here for years and it is a great place to hang out with friends 
	\\ and family. I love the food and service. I have never had a bad 
	experience when I am 
	there.\end{tabular}                                                         
	                         \\
	 \hline
	0.7                  & \begin{tabular}[c]{@{}l@{}}My family and I are huge 
	fans of this place. The staff is super nice and the food is great. The \\ 
	chicken is very good and the garlic sauce is perfect. Ice cream topped with 
	fruit is delicious \\ too. Highly 
	recommended!\end{tabular}                                                   
	    \\
	 \hline
	1.0                  & \begin{tabular}[c]{@{}l@{}}
    I had the grilled veggie burger with fries!!!!
        Ohhhh and taste. Omgggg! Very flavorful! It was \\so delicious that I didn't spell it!!
    \end{tabular}                                                         
	 \\
	 \hline
\end{tabular}
 \vspace{0.2in}
\caption{Example of the generated five-star fake reviews targeting at a buffet 
restaurant in NYC. More examples of
generated reviews are included in the Appendix~\ref{appendix:mg_example}.}
\label{tab:review_example}
 \vspace{-0.15in}
\end{table*}

%% file: eval.tex
\section{Evaluating Quality of Machine-Generated Reviews}
In this section, we evaluate the quality of machine-generated 
reviews along two dimensions. First, we investigate whether generated reviews 
can bypass detection by existing algorithmic approaches. Second, we conduct 
an end-to-end user study, by presenting restaurant reviews containing both 
generated reviews and real reviews to human judges. Our goal is to 
understand whether humans can distinguish generated reviews from real reviews.

\label{sec:eval}

\input{eval-mach}

\input{eval-user}

%% file: eval-mach.tex
\begin{figure*}[t]
	\begin{minipage}{.485\textwidth}
		\centering
		\includegraphics[width=\linewidth]{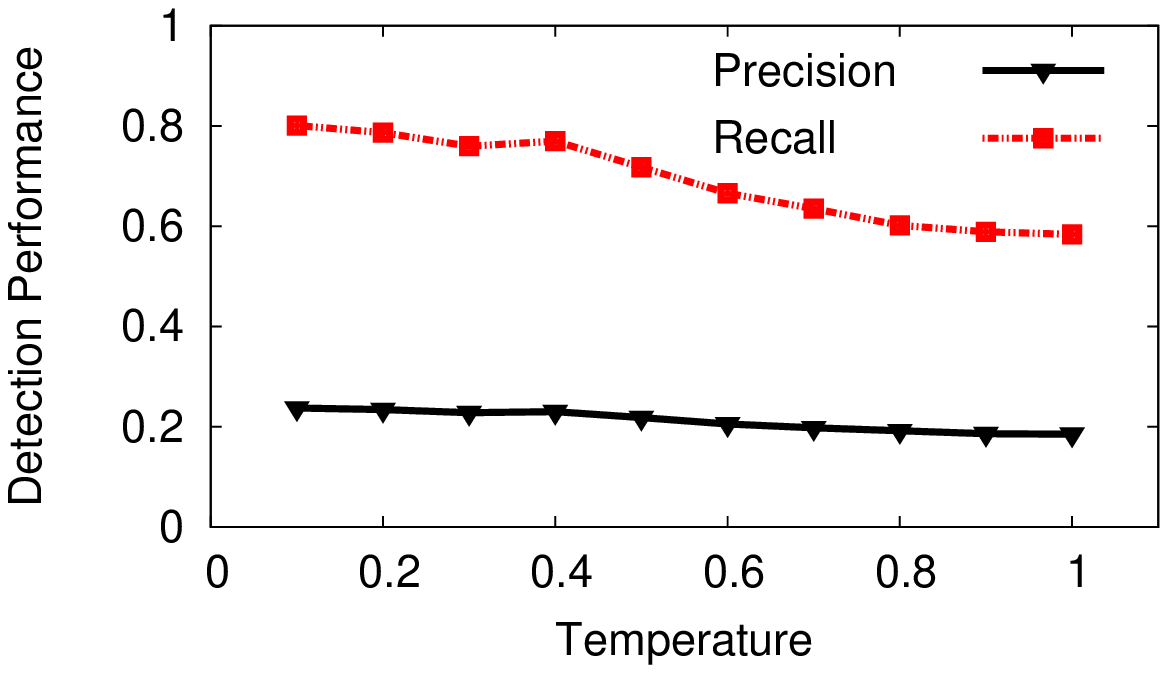}
		\caption{Performance of linguistic classifier on detecting 
			machine-generated reviews.}
		\label{fig:mch_test}
	\end{minipage}
	\hfill
	\begin{minipage}{.485\textwidth}
		\centering
		\includegraphics[width=\linewidth]{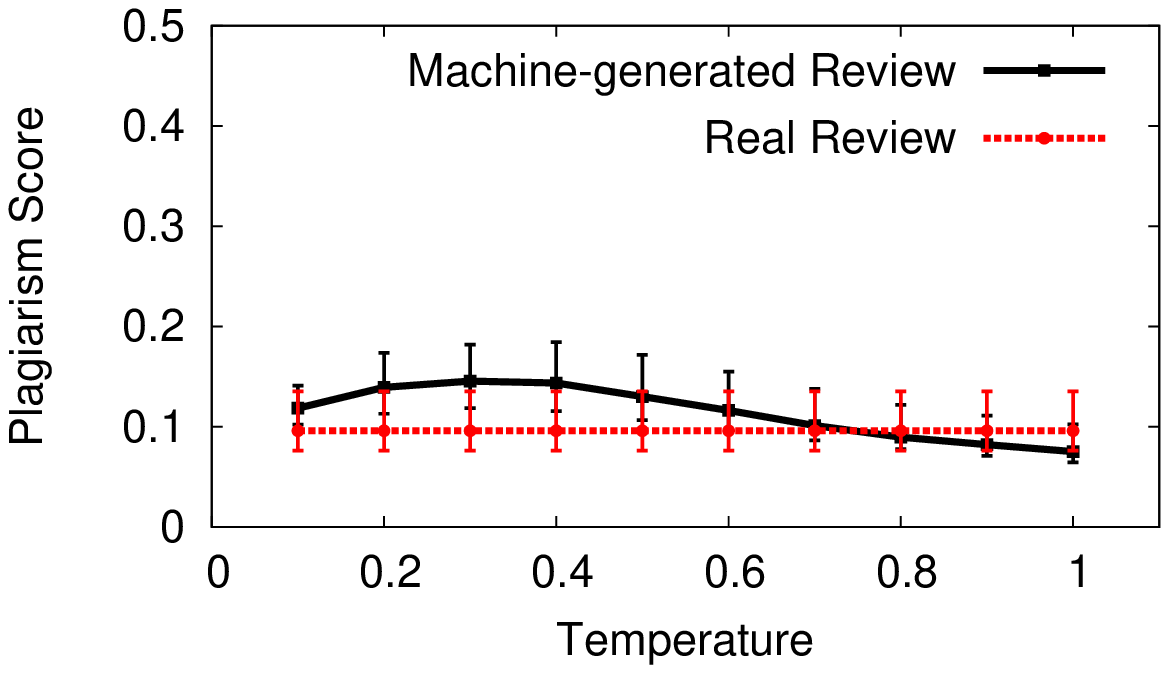}
		\caption{Plagiarism similarity score. Each point shows 
			median, 25th and 75th percentile of the score 
			distribution.}
		\label{fig:plag_result}
	\end{minipage}
\end{figure*}

\subsection{Detection by Existing Algorithms}
\label{subsec:mach_test}

We focus on two popular algorithmic techniques to distinguish machine-generated 
reviews 
from
real reviews: (1) a supervised ML scheme based on linguistic
features, (2) a plagiarism detector to  
check for duplications between machine-generated reviews and training set 
(real) reviews. 

\subsubsectitle{ML-based Review Filter.} Using machine learning classifiers
to detect fake reviews is a well studied problem~\cite{mukherjee2013yelp,
spam_1, spam_2}. 
Most of the prior
works rely on the observation that characteristics of fake reviews deviate from
real reviews along many linguistic dimensions. 
We identified 5 groups of linguistic features,
consisting of %
77 features total that %
previously demonstrated strong
discriminatory power %
for distinguishing fake and real reviews. We describe the features below:

\begin{packed_itemize}
	\item  \textit{Similarity feature (1)}: Captures inter-sentence similarity
	within a review at the word level. It is computed as the maximum cosine
	similarity between unigram features among all pairs of
	sentences~\cite{spam_4, spam_2, fei2013exploiting, yelp_data_2}.
   \item  \textit{Structural features (4)}: Captures the structural aspects of a
   review. Individual features include the number of words, the number of
   sentences, the average sentence length (\# of words) and the average word
   length (\# of characters)~\cite{spam_2, yelp_data_2}.
   \item  \textit{Syntactic features (6)}: Captures the linguistics properties
   of the review based on parts-of-speech (POS) tagging. Features include %
   (distinct) percentages of nouns, verbs, adjectives and adverbs, first personal pronouns, and 
   second personal pronouns~\cite{spam_4, spam_2, yelp_data_2}.
   \item  \textit{Semantic features (4)}: Captures the subjectivity and
   sentiment of the reviews. Features include percentage of subjective words,
   percentage of objective words, percentage of positive words and percentage of
   negative words. All these features are defined in
   SentiWordNet~\cite{sentiwordnet}, a popular lexical resource for opinion
   mining~\cite{spam_4, spam_1, yelp_data_2}.
   \item  \textit{LIWC features (62)}: The Linguistic Inquiry and Word Count
   (LIWC) software~\cite{liwc} is a widely used text analysis tool in the social
   sciences. It categorizes $\sim$4,500 keywords into $\sim$68 psychological
   classes (\eg linguistic processes, psychological processes, personal concerns
   and spoken categories). We use the percentage of word count in each class as
   a feature, and exclude the features already included in the previous
   groups~\cite{spam_6, mukherjee2013yelp}.
\end{packed_itemize}

We train a linear SVM classifier on the Yelp ground-truth dataset, composed of
real reviews (Yelp unfiltered reviews), and fake reviews (Yelp filtered 
reviews). %
After training with all 77 linguistic features, we tested the performance of the 
classifier on the Yelp attack dataset, 
composed of real reviews and machine-generated reviews. We run 10-fold cross 
validation and report the average performance.

\begin{figure*}[!t]
	\centering
	\mbox{
		\subfigure[Average word length (structural feature)]
		{\includegraphics[width=0.33\textwidth]{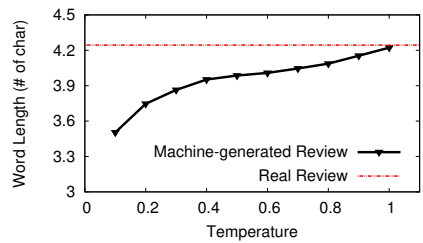}}
		\hspace{0.01in}
		\subfigure[Ratio of verb usage (syntactic feature)]
		{\includegraphics[width=0.33\textwidth]{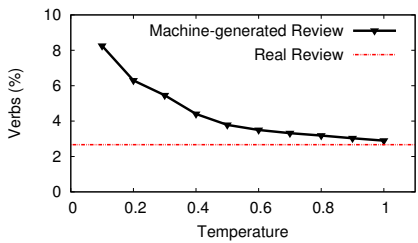}}
		\subfigure[Ratio of positive word usage (semantic feature)]
		{\includegraphics[width=0.33\textwidth]{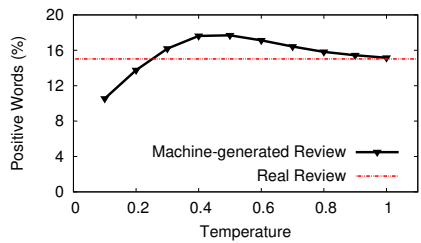}}
	}
	\caption{Change of linguistic feature values when temperature varies. }
	\label{fig:ling_ftr_demo}    
\end{figure*}

Evaluation of attack performance uses \textit{precision} (percentage of reviews 
flagged by the classifier that are fake reviews), and \textit{recall}
(percentage of fake reviews flagged by the classifier). %
Figure~\ref{fig:mch_test} shows the precision and recall of the classifier when
applied to machine-generated reviews generated from different temperatures 
(along with real reviews), 
with lower values indicating higher performing attacks. Overall,
we observe high performing attacks %
at all temperatures. 
The best attack is at
temperature 1.0, with a low precision of 18.48\%, and a recall of
58.37\%. Low precision indicates the inability of the ML classifier to 
distinguish between real reviews and machine-generated reviews.

In Figure~\ref{fig:mch_test}, we observe that attack performance increases
with temperature. To further understand this trend, we analyze how the
linguistic features of the generated text vary as we increase temperature. In
Figure~\ref{fig:ling_ftr_demo}, we compare the average value of a linguistic
feature of generated reviews with real reviews at different
temperatures. We show results for 3 linguistic features across 3 categories,
and other features exhibit similar trends.  In general, feature values of the
machine-generated reviews diverge from real reviews at low temperatures, 
and
converge as temperature increases, thus making it harder to distinguish them
from real reviews at high temperatures.

\subsubsectitle{Plagiarism Detector.} Achieving reasonable linguistic quality
does not rule out the possibility of being fake. A simple attack involves
generating fake reviews by duplicating or partially copying from real
reviews. In such cases, the review quality would be quite good, and would
pass the linguistic filter. Standard solution is to rely on plagiarism 
checkers to identify the duplicate or near-duplicate reviews. Given that the RNN
model is trained to generate text similar to the training set, we examine if the
machine-generated reviews are duplicates or near-duplicates of reviews in the
training set.

To conduct a plagiarism check, we assume that the service provider has access to
a database of reviews used for training the RNN. Next, given a machine-generated
review, the service provider runs a plagiarism check by comparing it with 
reviews in the
database. This is a best case scenario for a plagiarism test, and helps %
us understand its potential to detect generated reviews.

We use Winnowing~\cite{winnowing}, a widely used method to identify duplicate or
near-duplicate text. For a suspicious text, Winnowing first generates a set of
fingerprints by applying a hashing function to a set of substrings in the text,
and then compares the fingerprints between the suspicious text and the text in
database. Similarity between two reviews is computed using \textit{Jaccard
Similarity}~\cite{jaccard} of their fingerprints generated from Winnowing. The
plagiarism similarity score for a single review is computed as the max
similarity with all the other reviews in the dataset, and ranges from 0 to 1 (1
indicates identical reviews).

We pick a random sample of 10K machine-generated reviews for the plagiarism
test, and the database (for comparison) includes the entire Yelp training
dataset. Figure~\ref{fig:plag_result} shows the quantiles of similarity
scores at different temperatures. Each point shows median, 25th and 75th
percentile of the plagiarism score distribution. In addition, we also show
the similarity score distribution for real reviews, which serves
as a baseline for comparison. Note that scores for real reviews do
not vary with temperature. We obverse that plagiarism scores of
machine-generated reviews are low at all temperatures (lower score represents 
smaller probability of copying) and decrease as temperature increases.
In addition, machine-generated reviews and real reviews 
show similar plagiarism scores, thus making them harder to distinguish. 
For example, at temperature 1.0, if we set a plagiarism score threshold such
that 95\% of real reviews are not flagged, we observe that 96\% of
machine-generated reviews still bypass the check. Thus, it remains hard to
detect machine-generated reviews using a plagiarism checker without
inadvertently flagging a large number of real reviews. This shows that
the RNN does not simply copy the existing reviews from the training set.

%% file: eval-user.tex
\subsection{Evaluation by User Study}
\label{subsec:human_test} 

Regardless of how well machine-generated reviews perform on statistical 
measures and
tests, the real test is whether they can pass for real reviews when read by
human users. In this section, we conduct an end-to-end user study 
to evaluate whether human examination can detect machine-generated reviews. 
In practice, service providers are known to involve human content moderators
to separate machine-generated reviews from real reviews~\cite{yelp_human}. More
importantly, these tests will tell us how convincing these reviews are to
human readers, and whether they will accomplish their goals of 
manipulating user opinions.

\subsubsectitle{User Study to Detect Machine-Generated Reviews.} To
measure human performance, we conduct surveys\footnote{Prior to conducting our
study, we submitted a human subject protocol and received approval from our
local IRB board.} on Amazon Mechanical Turk
(AMT\footnote{\url{https://www.mturk.com/}}). Each survey includes a restaurant
name, description (explaining the restaurant category and description provided
by the business on Yelp), and a set of reviews, which includes machine-generated
reviews and real reviews written for that restaurant. We then ask
each worker to mark reviews they consider to be fake, using any basis for their
judgment.

For our survey, we choose 40 restaurants with the most number of reviews in
our ground-truth dataset. For each restaurant, we generate surveys, each of
which include 20 random reviews, 
out of which some portion ($X$) are machine-generated reviews, and the
rest are real reviews from Yelp.  The number $X$ is randomly selected between
0 to 5 so that the expected ratio of fake reviews (12.5\%) matches the real
world setting (11.99\% in Table~\ref{tab:eval_datasets}). Additionally, we
control the quality of real reviews shown in the surveys to cover the full
range of usefulness. We leverage the \textit{review usefulness} (a simple count
of the number of users who found the review to be useful) metadata provided
by Yelp for each review.

For each of our 40 restaurants, we generated reviews using 5 different
temperature parameters: [0.1, 0.3, 0.5, 0.7, 1.0]. We give each unique survey
to 3 different workers, giving us a total of 600 surveys. Out of these 600
responses, we discarded 6 because they did not mark the gold standard
reviews. Gold standard reviews are basically strings of random characters
(\ie meaningless text), that looks clearly fake to any worker. Lastly, we
only request \textit{master
  workers}\footnote{\url{https://www.mturk.com/mturk/help?helpPage=worker\#what_is_master_worker}}
located in the US to guarantee English literacy. We show an example of our
survey in the Figure~\ref{fig:suverys}(a) in Appendix~\ref{appendix:survery}.

Figure~\ref{fig:human_perf} shows the human performance results as we vary
the temperature. First, we observe that machine-generated reviews appear
quite robust against a human test. Under the best configuration, the
precision is only 40.6\% with a recall of 16.2\%. In addition, similar to
algorithmic detection, attack performance improves as temperature
increases. This is surprising, since we would expect that reviews at the
extreme high or low temperature parameters would be easily flagged (either
too repetitive or too many grammatical/spelling errors). We saw earlier that
higher temperature produced reviews more statistically similar to real
reviews, but expected errors to make those reviews detectable by
humans. Instead, it seems that human users are much more sensitive to
repetitive errors than they are to small spelling or grammar mistakes.  We do
observe that the best attack performance occurs at a high temperature of 0.7,
which is marginally better than the performance at temperature of
1.0.

\subsubsectitle{Helpfulness of Machine-Generated Reviews.} Previously, we showed
that humans tend to mark many machine-generated reviews as real. This raises
a secondary question: \textit{For machine-generated reviews that are not caught
by humans, do they still have sufficient quality to be considered useful by a
user?} Answering this question, takes us a step further towards generating 
highly deceptive fake reviews. 
We run a second round of AMT surveys to investigate this question.

In each survey, we first asked the workers to mark reviews as fake or real.
Additionally, for the reviews marked as real, we asked for a rating of 
the usefulness of the review on a scale from 1 to 5 (1 as least
useful, 5 as most useful). An example of the survey is shown in the
Figure~\ref{fig:suverys}(b) in Appendix~\ref{appendix:survery}. We conduct the
survey using reviews generated at a temperature of 0.7, which gave the best
performance from the previous round. Also, in this round, we test on 80
restaurants and hire 5 workers for each restaurant. The rest of the survey
configuration remains the same as the first round.

We received all 400 responses and discarded 5 of them 
for failing the gold standard review. 
The average usefulness score of false negatives (unflagged machine-generated
reviews) is close to that of true negatives (unflagged Yelp real reviews):
machine-generated reviews have an average usefulness score of 3.15, which is
close to the average usefulness score of 3.28 for real Yelp reviews. That is
to say, workers think of unflagged machine generated reviews almost as 
useful as real reviews.

Overall, our experiments find machine-generated reviews very close to
mimicking the quality of real reviews. Furthermore, the attacker is
incentivized to generate reviews at high temperatures, as such reviews appear
more effective at deceiving users.

\begin{figure}[t]
	\centering
	\includegraphics[width=\linewidth]{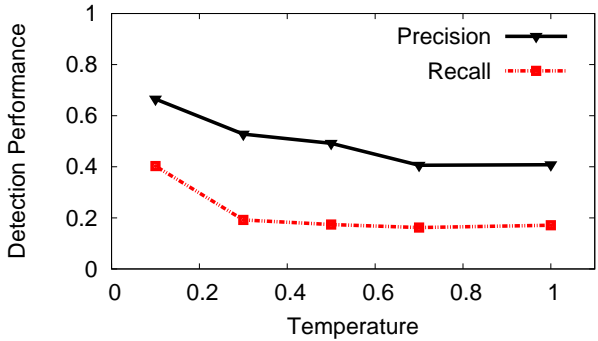}
	\caption{Performance of human judgment on detecting machine-generated 
		review.}
	\label{fig:human_perf}
\end{figure}

%% file: defense.tex
\section{Defending Against Machine-generated reviews}
\label{sec:defense}

\input{defense-overview}

\input{defense-eval}

%% file: defense-overview.tex
In this section, we propose a supervised learning scheme to detect
machine-generated reviews with high accuracy.

\subsubsectitle{Assumption. } We assume that the service provider has access to 
a
limited set of reviews generated by an attacker's language
model and a set of real reviews available on the review site.

\subsubsectitle{Why is defense challenging?} Fundamentally, we are trying to
develop a machine learning classifier capable of detecting the output of
another ML model. In an ideal scenario, this seems impossible, because the
generator model and detector are likely using the exactly same metrics on the
same inputs (or training datasets).  Thus, any metric that an ML
detector is capable of detecting can be accounted for by the ML-based generator.

\begin{figure}[!t]
	\centering 	
	\includegraphics[width=0.465\textwidth]{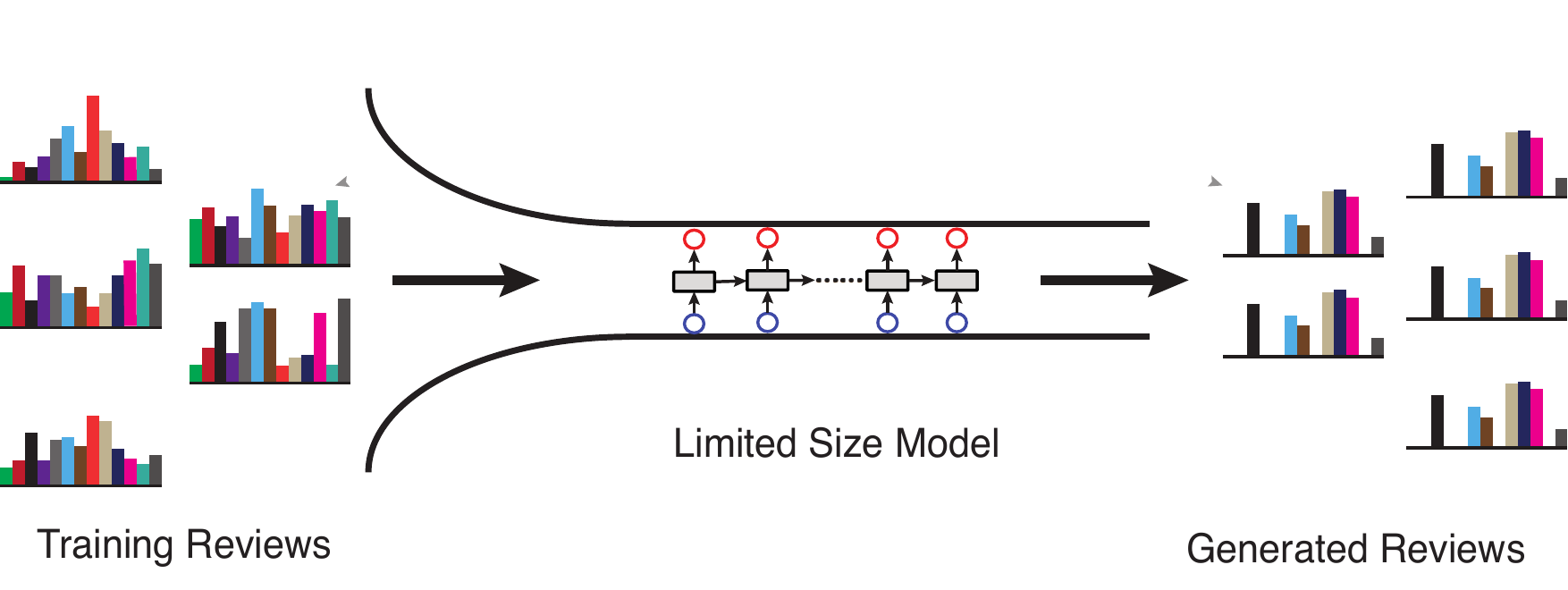}
	\caption{Key insight of our defense. During the training process, 
		the language model builds a fixed memory representation of the large 
		training 
		corpus and its representativity is limited by the model size. The 
		information loss incurred during the training would propagate to 
		the generated text, leading to statistically detectable difference in 
		the
		underlying character distribution between the generated text and human 
		text.}
	\label{fig:def_demo}
\end{figure}

\subsubsectitle{Key Insight.} Figure~\ref{fig:def_demo} shows the key
intuition behind our defense. While we expect that an attacker is aware of
any metric used by the ML detector for detection, we can find a sufficiently
``complex'' metric where accurately capturing (and reproducing) the metric
requires an extremely large RNN infrastructure that is beyond the means of
most attackers.  This leverages the fact that a generative language model
builds a fixed memory representation of the entire training corpus, which
limits the amount of information that can be learned from a training corpus.
More specifically, we observe that text produced naturally (\eg by a human)
diverges from machine generated text when we compare the \textit{character
  level distribution}, even when higher level linguistic features (\eg
syntactic, semantic features) might be similar. Such divergence in the
character level distribution is primarily due to the information loss
inherent in the machine-generation process, which is a function of the
modeling power of the RNN used.

We note that the character level distribution metric is appropriate for our
purposes, because it is the lowest level metric being modeled by our RNN, and
therefore most likely to become a complexity bottleneck for the RNN. An
attacker might consider an RNN generator that trains using word-level
distributions. Fortunately (for our purposes), word-level distributions are
more complex and difficult to model, both because the number of words is
combinatorially larger than number of valid characters in the English
language, and because such a model would need to add additional rules for valid
punctuation. Therefore, an RNN generator targeting word-level
distributions would be even more computationally constrained (thus incur more 
information loss), and
intuitively, our defense would work at least as well as on a character-level
RNN. 

\subsection{Proposed Methodology}
More concretely, consider the following attack scenario: The attacker trains
on a set of human generated reviews $\text{R}_\text{T}$ and builds a
character-level RNN language model M, to generate a set of reviews
$\text{R}_\text{F}$.
Even when $\text{R}_\text{T}$ is chosen in such a way that
$\text{R}_\text{F}$ becomes linguistically similar to the real reviews
$\text{R}_\text{L}$ on the site, we can statistically detect variations in
the character level distribution between $\text{R}_\text{F}$ and
$\text{R}_\text{L}$. 
Algorithm~\ref{alg:defense} provides details of the method. The service provider
maintains access to the set of known machine-generated reviews
$\text{R}_\text{F}$, along with the set of real reviews
$\text{R}_\text{L}$, and aims to determine whether a given test review T is
fake or real.
Based on our insight, we expect the character-level distribution of reviews
in set $\text{R}_\text{F}$ to statistically diverge from that of reviews
in set $\text{R}_\text{L}$.

The character-level probability distribution
$P(\text{X}_{\text{t+1}}$=$\text{x}_{\text{t+1}}$|$\text{x}_{{\text{1},\dots,\text{t}}})$,
gives the probability of predicting the next character, given the sequence of
preceding characters. To capture the divergence in the character
distribution, the defender first builds an RNN language model
$\text{RNN}_\text{F}$ using the set of machine-generated reviews
$\text{R}_\text{F}$ and another language model $\text{RNN}_\text{L}$ using
$\text{R}_\text{L}$.
Next, given a test review $T$, we feed the review, character by character, into
each RNN model to obtain two character distributions, providing statistical
representations of review $T$ for each model.  Finally, if the
log-likelihood ratio of the test review's character distribution closely fits
the model $\text{RNN}_\text{F}$, then we flag the review as fake.

\begin{algorithm}[!t]
	\caption{Proposed Defense}\label{alg:defense}
	\begin{algorithmic}[1]
		\LeftComment{input-$\text{R}_\text{F}$:machine-generated review 
		training set, 
		$\text{R}_\text{L}$:real
			review training set, T:test review}
		\Procedure{Defense}{$\text{R}_\text{F}$, $\text{R}_\text{L}$, T}
		\State N $\leftarrow$ length(T)
		\State $\text{RNN}_\text{F}$ $\leftarrow$ 
		trainRNN($\text{R}_\text{F}$)
		\State $\text{RNN}_\text{L}$ $\leftarrow$ 
		trainRNN($\text{R}_\text{L}$)
		\For{t = 1:N-1}
		\State feed $\text{X}_\text{t}$ into $\text{RNN}_\text{F}$
		\State $\mathcal{L}_\text{F}$ $\leftarrow$ 
		$\text{P}_\text{F}$($\text{X}_\text{t+1}$=$\text{x}_\text{t+1}$|$\text{x}_{\text{1},\dots,\text{t}}$)
		\State feed $\text{X}_\text{t}$ into $\text{RNN}_\text{L}$
		\State $\mathcal{L}_\text{L}$ $\leftarrow$ 
		$\text{P}_\text{L}$($\text{X}_\text{t+1}$=$\text{x}_\text{t+1}$|$\text{x}_{\text{1},\dots,\text{t}}$)
		\State $\mathcal{L}_\text{t}$ $\leftarrow$
		$-\log{\frac{\mathcal{L}_{\text{L}}}{\mathcal{L}_{\text{F}}}}$
		\Comment{negative log-likelihood ratio}
		\EndFor
		\State $\bar{\mathcal{L}}$ $\leftarrow$
		$\frac{\sum_{\text{i=1}}^{\text{N-1}}{\mathcal{L}_{\text{i}}}}{\text{N-1}}$
		\If{$\bar{\mathcal{L}} > 0$}
		\State return FAKE
		\Else
		\State return REAL
		\EndIf

		\EndProcedure
	\end{algorithmic}
\end{algorithm}

%% file: defense-eval.tex
\subsection{Defense Evaluation}
\label{subsec:def_eval}

\begin{figure*}[!t]
	\begin{minipage}{0.615\textwidth}
		\vspace{-0.3in}
		\centering 		\mbox{    		 \subfigure[Precision]
			{\includegraphics[width=0.49\textwidth]{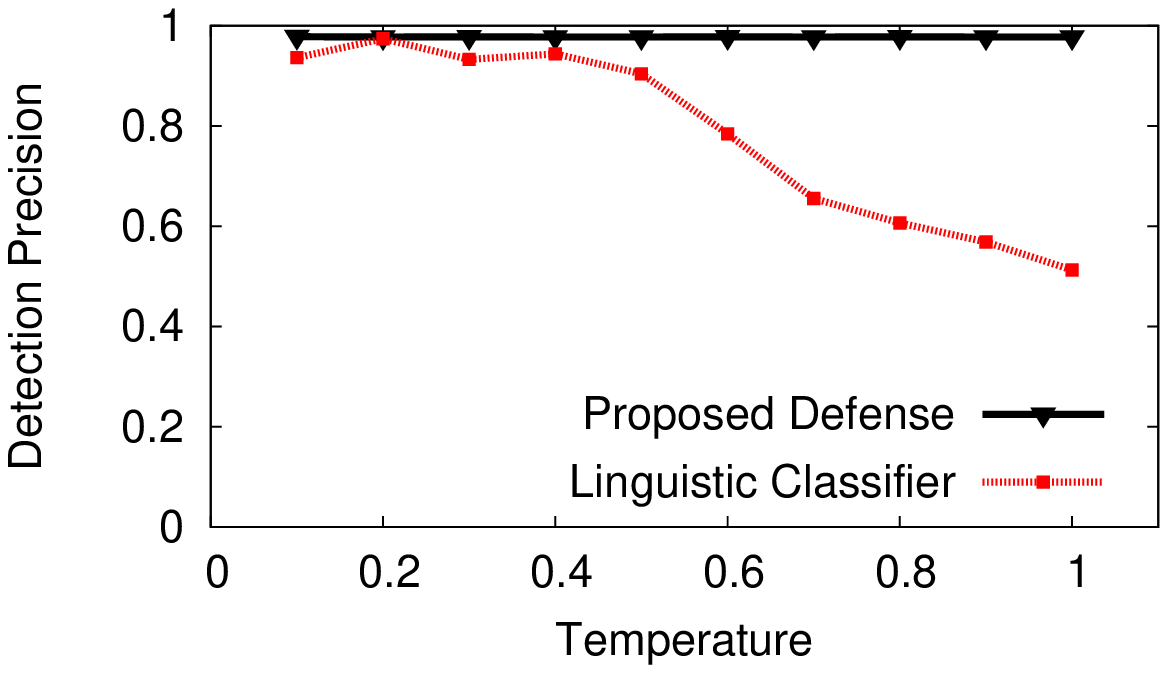}}
			\hspace{0.01in}     		\subfigure[Recall]
			{\includegraphics[width=0.49\textwidth]{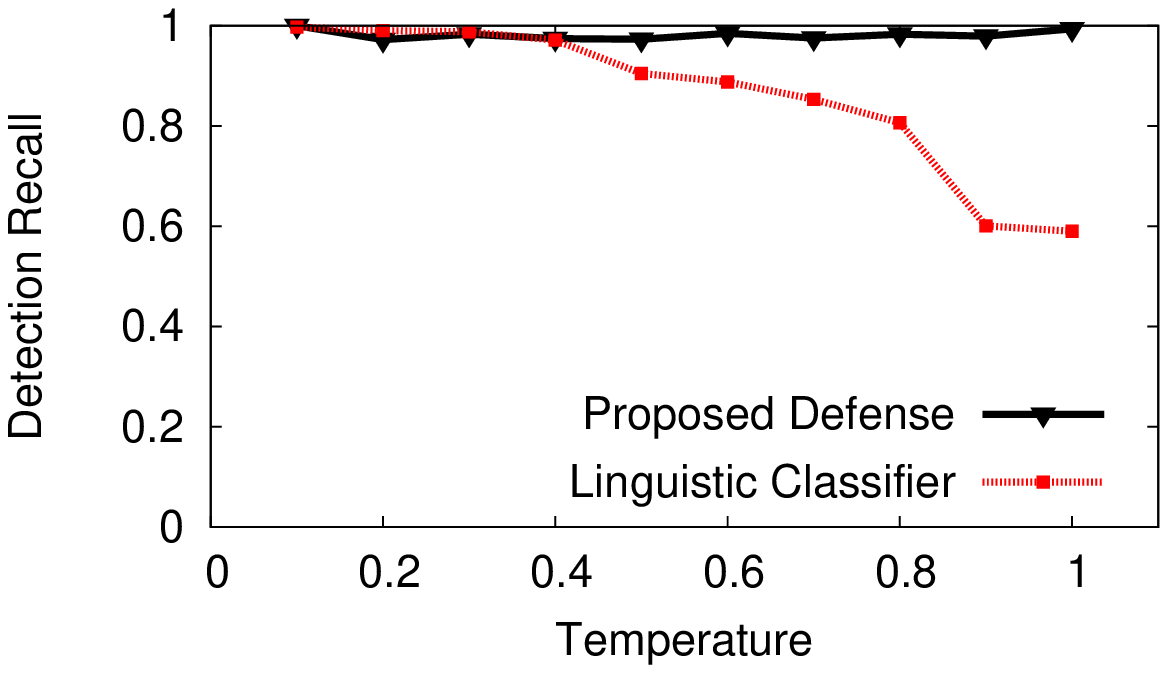}}
			} 		\caption{Performance of proposed defense and linguistic
            classifier (Section~\ref{subsec:mach_test}).}
		\label{fig:adv_perf}        
	\end{minipage}
	\hfill   
	\begin{minipage}{.335\textwidth}
		\centering
		\includegraphics[width=\linewidth]{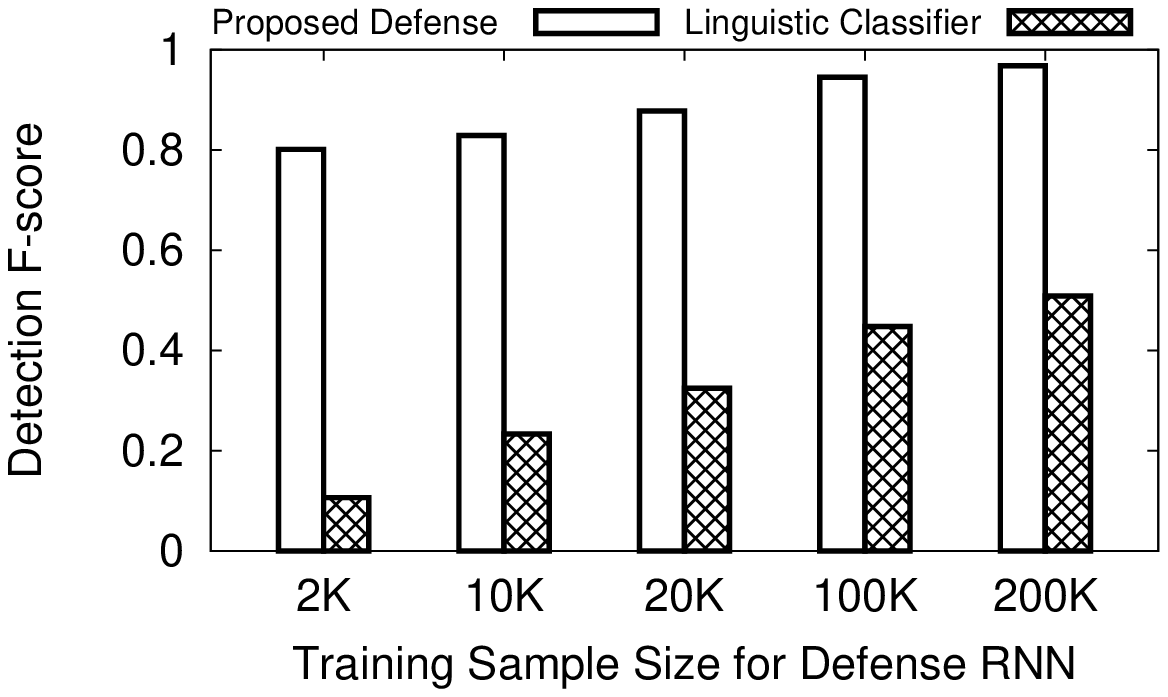}
		\vspace{-0.1in}
		\caption{Detection performance given different sizes of defense training
			samples.}
		\label{fig:def_vary_size}
	\end{minipage}
\end{figure*}

We evaluate our defense scheme along two directions. First, we study the
detection performance of our scheme and how it compares to an ML scheme based
on linguistic features (Section~\ref{subsec:mach_test}). Second, we
investigate the robustness of our approach to evade detection against two
attacker strategies.

We follow a standard RNN model training process to train $\text{RNN}_\text{F}$ 
and $\text{RNN}_\text{L}$. We refer to them as \textit{defense RNN}. 
Unless otherwise stated, we report performance on 2,000 test reviews 
(balanced set of real and machine-generated reviews) in the remainder of 
this section.

\subsubsectitle{Detection Performance. } To first understand the performance
in a potential best case scenario, we evaluate our scheme by considering a
large amount of ground-truth information. Our ground-truth set consists of
120K machine-generated reviews, from our Yelp attack dataset
(Section~\ref{subsec:data}), and 120K additional real reviews. We set
the model configuration for the defense RNN to be the following: 1,024 hidden
units, 2 hidden layers, batch size of 128 and 20 training epochs.

As a baseline for comparison, we compute detection performance using
the ML scheme described in Section~\ref{subsec:mach_test}, which we refer to as
the \textit{linguistic classifier}. Note that the ML scheme is based on high
level linguistic features and trained using the same ground-truth set of
machine-generated and real reviews. We expect the linguistic classifier to
perform better than the results in Section~\ref{subsec:mach_test} because the
training data now includes machine-generated reviews that we aim to identify 
directly. 

Figure~\ref{fig:adv_perf} shows the detection performance when we train and test
on text generated at different temperatures. Our approach achieves high
precision and recall at all temperatures, \ie over 0.98 precision and 0.97 
recall. Additionally, we outperform the linguistic classifier at most
temperatures, and the gap between the two schemes increases at higher
temperatures (\eg temperature > 0.6). At temperature 1.0, our scheme achieves an
\textit{F-score} (the harmonic mean of precision and recall) of 0.98, 
while the linguistic approach only achieves an F-score of
0.55. Interestingly, the linguistic classifier shows high detection performance
at low temperatures. This trend can be explained by our earlier finding that
linguistic features diverge more from the real reviews at low temperatures
(Figure~\ref{fig:ling_ftr_demo}).

\begin{table}[!b]
	\begin{minipage}{.485\textwidth}
	\centering
	\begin{tabular}{@{}ccccc@{}}
		\toprule
		\begin{tabular}[c]{@{}c@{}}Training\\ Samples\end{tabular} & 
		\begin{tabular}[c]{@{}c@{}}Hidden\\  Unit Size\end{tabular} & 
		\begin{tabular}[c]{@{}c@{}}Layer \\ Size\end{tabular} & 
		\begin{tabular}[c]{@{}c@{}}Batch\\ Size\end{tabular} & 
		\begin{tabular}[c]{@{}c@{}}Training\\ Epoch\end{tabular} \\ \midrule
		2K                                                        & 
		128                                                         & 
		1                                                     & 
		16                                                   & 
		50                                                       \\
		10K                                                       & 
		256                                                         & 
		1                                                     & 
		32                                                   & 
		50                                                       \\
		20K                                                       & 
		512                                                         & 
		1                                                     & 
		56                                                   & 
		30                                                       \\
		100K                                                      & 
		768                                                         & 
		2                                                     & 
		128                                                  & 
		20                                                       \\
		200K                                                      & 
		1,024                                                        & 
		2                                                     & 
		128                                                  & 
		20                                                       \\ \bottomrule
	\end{tabular}	
	\vspace{0.2in}
	\caption{Training configurations of defense models when defense training 
		sample size varies.}
	\label{tab:def_config}
	\end{minipage}
	\vfill
	\begin{minipage}{.485\textwidth}
	\centering
	\begin{tabular}{@{}ccccc@{}}
		\toprule
		\begin{tabular}[c]{@{}c@{}}Hidden\\ Unit Size\end{tabular} & 
		\begin{tabular}[c]{@{}c@{}}Training \\ Samples\end{tabular} & 
		\begin{tabular}[c]{@{}c@{}}Layer \\ Size\end{tabular} & 
		\begin{tabular}[c]{@{}c@{}}Batch\\ Size\end{tabular} & 
		\begin{tabular}[c]{@{}c@{}}Training\\ Epoch\end{tabular} \\ \midrule
		128                                                        & 
		10K                                                         & 
		1                                                     & 
		32                                                   & 
		50                                                       \\
		256                                                        & 
		50K                                                         & 
		1                                                     & 
		56                                                   & 
		50                                                       \\
		512                                                        & 
		100K                                                        & 
		1                                                     & 
		128                                                   & 
		30                                                       \\
		768                                                        & 
		500K                                                        & 
		2                                                     & 
		256                                                  & 
		20                                                       \\
		1,024                                                       & 
		617K                                                        & 
		2                                                     & 
		256                                                  & 
		20                                                       \\
		2,048                                                       & 
		617K                                                        & 
		2                                                     & 
		256                                                  & 
		50                                                
		       \\ 
		\bottomrule
	\end{tabular}
	\vspace{0.2in}
	\caption{Training configurations of attack models when attack model size 
		varies.}
	\label{tab:model_size}
\end{minipage}
\end{table}

Next, we study performance when we limit the amount of ground-truth used for
training and focus on text generated at temperature 1.0.
Figure~\ref{fig:def_vary_size} shows performance when training data size
varies from 2,000 to 200K samples. Each training dataset is a balanced dataset
of machine-generated and real reviews. The RNN model configuration used for
defense for each training set size is detailed in
Table~\ref{tab:def_config}. Our scheme significantly outperforms the
linguistic classifier for all datasets and achieves a high F-score of 0.80
using only 1,000 machine-generated reviews (2,000 training
dataset). Considering the fact that service providers have taken considerable
effort to build large fake review datasets ($\sim$250K fake reviews in
Table~\ref{tab:eval_datasets}), 1,000 reviews is a relatively small sample of
known fake reviews.  Thus, unlike the linguistic classifier, our defense
scheme performs well with highly limited ground-truth information.

\begin{figure*}[t]
	\begin{minipage}{.49\textwidth}
		\centering
		\includegraphics[width=\linewidth]{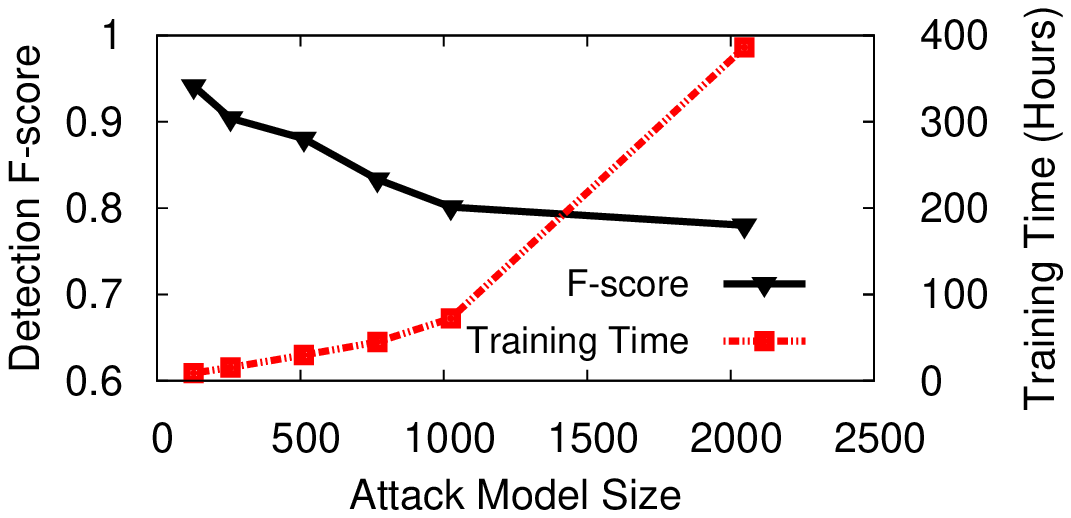}
		\caption{Detection performance against attacks generated by different 
		models and their training costs.}
		\label{fig:model_vs_perf}
	\end{minipage}
	\hfill
	\begin{minipage}{.49\textwidth}
		\centering
	\includegraphics[width=\linewidth]{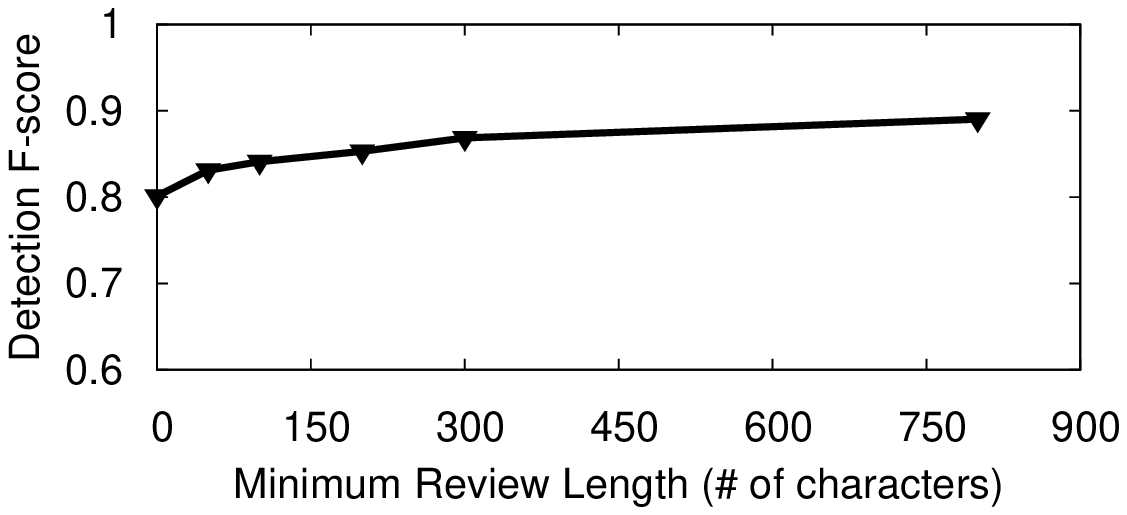}
		\caption{Detection performance when the minimum review length is 
		restricted.}
		\label{fig:attack_cost}
	\end{minipage}
\end{figure*}

\subsubsectitle{Evading Detection by Increasing Attack Model Quality. }  The
attacker can evade detection by improving the quality of the RNN generative
model or increasing the memory size of the model. A higher quality model
would generate more ``natural'' text, and thus reduce the divergence in the
character distribution. However, this strategy comes with a higher cost for
the attack. Training a larger RNN model requires more training data to be
collected, more computational resources (\eg GPUs with more memory and
computational capacity), while also imposing additional training time. It is
hard to quantify the increase in attack cost when accounting for all these
factors. Instead, we focus on the impact on training time as the attacker
varies model quality to evade detection.

We vary the attack model size (number of hidden units) from 128 to 2,048. We
also vary other model parameters, and the size of the training dataset to avoid
underfitting or overfitting. Details of the attack models are described in
Table~\ref{tab:model_size}. For the defense RNN, we use a
configuration based on 2K training samples in Table~\ref{tab:def_config}. 

Figure~\ref{fig:model_vs_perf} shows the tradeoff between decreases in detection
performance (F-score) and increases in training time for the attacker. In
general, when the attacker trains a larger model, our defense performance would
degrade: when doubling the model size from 128 to 256 cells, detection
performance drops by 3.95\% with training time increasing by 71.86\%. 
As the model size grows further, attacker's gain in evasion rate 
slows, while the
matching training time accelerates significantly. This is due 
to the increase in computational complexity: from 
model size 1,024 to 2,048, defense performance only decreases by
2.70\% but the attacker's training time raises by 435.1\%.

In practice, larger models would require a significantly larger training set
as well. For example, Jozefowicz \etal~\cite{lm_limit} trained a 2-layer RNN
with $8,192$ hidden units on a dataset with $\sim$$0.8B$ words, 
14x larger than our training dataset, to achieve state-of-the-art model
performance. Therefore, the computational cost and amount of training data
required to train a larger model would become prohibitively expensive for all
but the most resourceful attackers.

Next, we show there are other ways to further diminish the power of any
resource-based countermeasures by the attacker.  We observe that our defense
scheme performs better on longer reviews, as the scheme has more data to
capture divergence in the character distribution. Based on this observation,
we propose a simple policy for the service provider to further raise the bar for
evasion: set a minimum review length. Figure~\ref{fig:attack_cost} shows how
detection performance varies as we increase the minimum review length
requirement against an attack model with $1,024$ hidden units.  Note that the
average review length on Yelp Training Data (Section~\ref{subsec:data}) is
483 characters. When we increase the minimum length requirement to 300
characters (\ie still below the average), F-score increases from 0.80 to
0.86, which nullifies the attack success gained by increasing model size from
512 to 1,024 hidden units (from Figure~\ref{fig:model_vs_perf}). This means
that the attacker now must aim for training a significantly larger attack
model, at the expense of increased training cost, to overcome the reduction
in attack success.

\begin{figure}[t]
	\centering 		\mbox{    		 \subfigure[Linguistic classifier]
		{\includegraphics[width=0.235\textwidth]{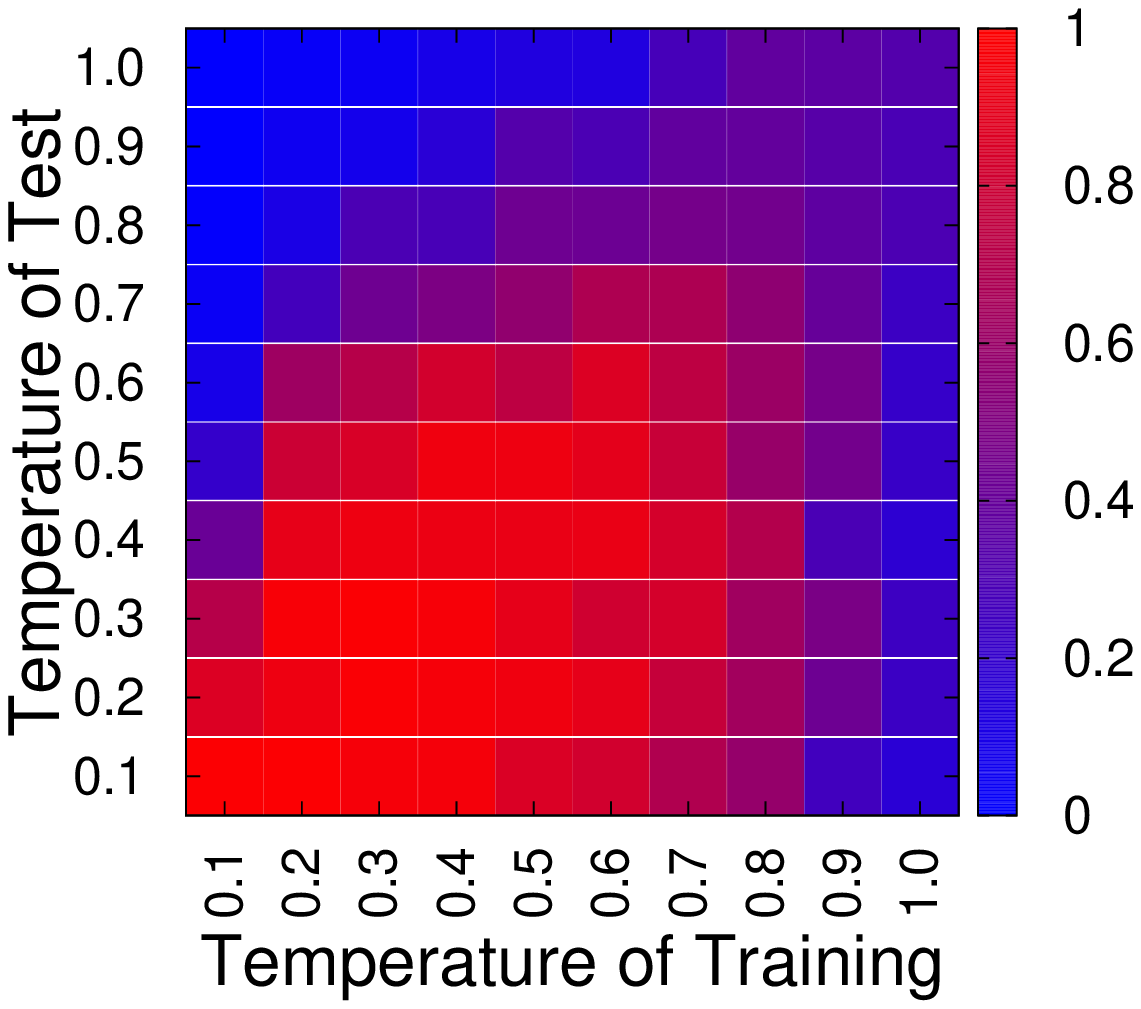}}
		\hspace{0.01in}     		\subfigure[Our method]
		{\includegraphics[width=0.235\textwidth]{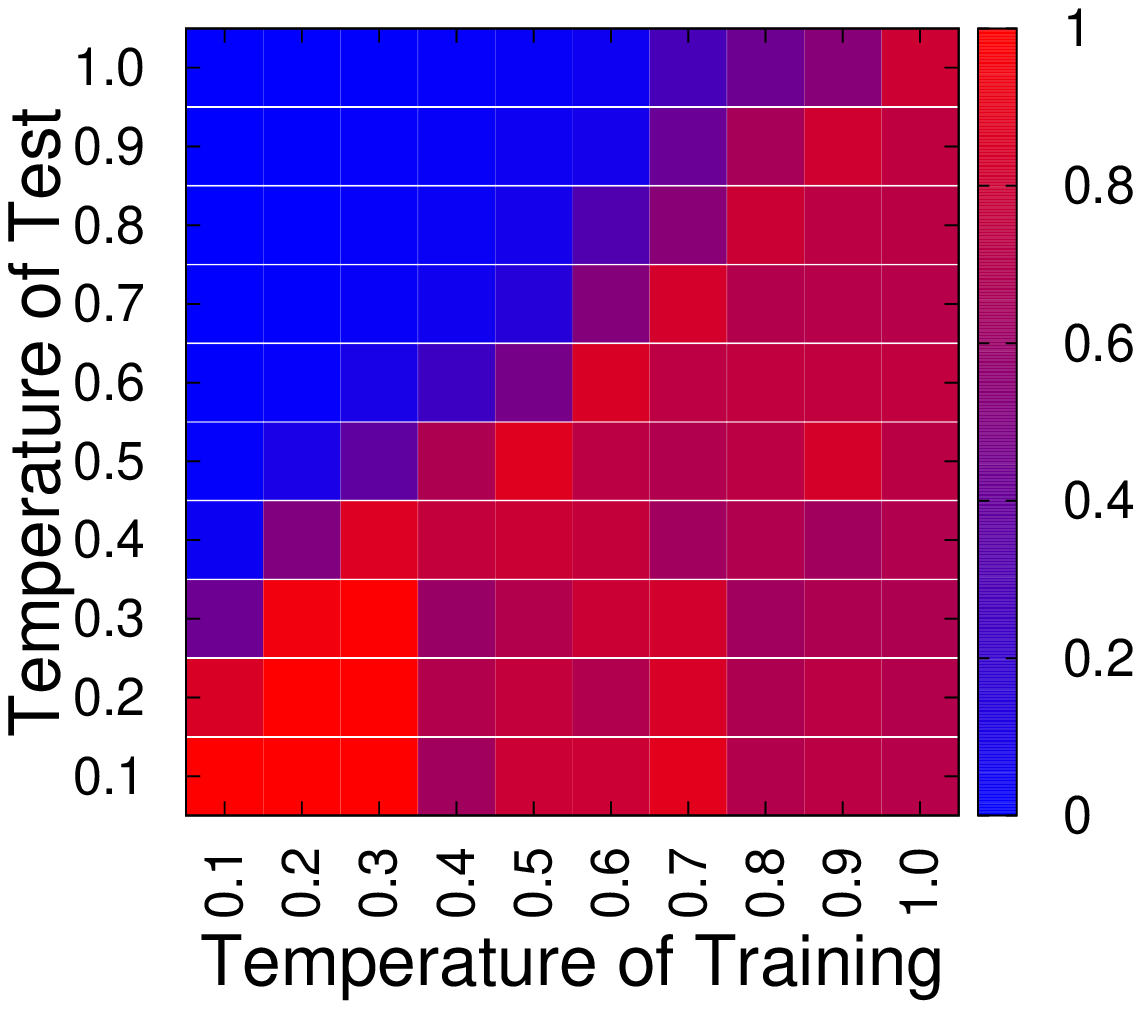}}
	} 	\caption{Detection performance (F-score) of training and then applying 
	at
		different temperatures.}
	\label{fig:perf_heatmap}
\end{figure}

\subsubsectitle{Evading Detection by Generating Reviews at Different
Temperatures. } When reviews from a language model are detected as fake,
one would expect the attacker to build a new language model (thus raising the cost) 
for the next attack. Instead, the attacker can try to evade further detection 
by generating new reviews using the existing model, by changing the temperature 
parameter (without re-training).  While our scheme can detect reviews at 
different temperatures when we have ground-truth information at those 
temperatures (Figure~\ref{fig:adv_perf}), performance remains unclear when 
ground-truth information is unavailable. Put differently: \textit{Using a 
defense scheme trained on reviews at a specific temperature, can we detect 
reviews generated at other temperatures?} If this is possible, it would allow 
the service provider to defend against such attacks without having to collect new 
ground-truth information.

We investigate the above scenario in Figure~\ref{fig:perf_heatmap}, where we 
train the linguistic classifier and our scheme at a specific temperature 
($\text{T}_\text{train}$) 
and evaluate detection performance at all other temperatures ($\text{T}_\text{test}$). 
Both $\text{T}_\text{train}$ and $\text{T}_\text{test}$ are from 0.1 to 1.
The training configuration is same as the one used for 2K training
samples in Table~\ref{tab:def_config}. 

As expected, the linguistic classifier exhibits poor defensive power at
higher temperatures. At low temperatures, we see that a defense trained at a
given temperature maintains effectiveness at temperatures in the
``neighborhood'' of that region, possibly due to similar linguistic
characteristics in the temperature neighborhood.  However, the performance
drops quickly when $\text{T}_\text{train}$ and $\text{T}_\text{test}$ are
far apart.

On the other hand, our approach shows an interesting trend. Unlike the
linguistic classifier, our performance maintains robustness whenever
$\text{T}_\text{train}$ $>$ $\text{T}_\text{test}$. This is because a review
generated at a high temperature would include both infrequent and frequent
patterns from the training sequence. As a result, a defense trained at a
higher temperature can capture the frequent sequences present in the
character distribution of a review at a lower temperature. Hence, our defense
scheme maintains high performance even when $\text{T}_\text{train}$ and
$\text{T}_\text{test}$ are distant as long as $\text{T}_\text{train}$ $\geq$
$\text{T}_\text{test}$. It should be noted that the attacker has an incentive
to generate reviews at high temperatures because they are more likely to
deceive users (Section~\ref{subsec:human_test}).  As such, the service provider 
can likely obtain some initial
ground-truth information about reviews at high temperatures, and thus build a
robust defense.

%% file: related.tex
\section{Related Work}
\label{sec:related}

\subsubsectitle{Text Generation.} There are a number of natural language 
generation techniques %
based on pre-defined templates~\cite{nlg_book,nlg_trad_2,nlg_trad_4,sun2013synthetic}.
These systems 
usually require %
some domain knowledge and well-designed 
rules. %
Recently, learning-based approaches became %
popular, \ie building a statistical 
model to learn the language information from a large corpus. %
In these cases, the quality of the generated text highly correlates with the 
model quality. Previous work shows that well-trained RNN models outperform
simpler language models like N-gram~\cite{lm_limit} and RNN-based language models have 
appeared as a promising approach to generate text~\cite{gen_rnn1, 
graves2013generating}.
Researchers have also achieved 
successful 
results 
in generating text for different domains,
including email responses~\cite{rnn_app_1}, image
description~\cite{rnn_image_desp},
movie dialogues~\cite{rnn_app_2} and online social network
conversations~\cite{rnn_app_3}.

Our work is closest to studies by Lipton \etal~\cite{lipton2015capturing},
Hovy~\cite{fake_review_user_study} and Lappas~\cite{adverial_2}.  Lipton
\etal~\cite{lipton2015capturing} also studies product review generation using
an RNN, but does not consider an adversarial setting.
Hovy~\cite{fake_review_user_study} performed a preliminary investigation of
n-gram-based review generation in an adversarial setting. Unlike our
research, they do not consider more sophisticated generative models (RNN) to
evade detection, and do not consider any robust defenses or countermeasures.
Lastly, Lappas~\cite{adverial_2} analyzes fake reviews from the attacker's
perspective to determine the factors that enable a successful attack, but does
not consider automatic generation of reviews.

\subsubsectitle{Neural Networks and Security.} 
Most prior studies on applying
Neural Networks focus on improving existing defenses against various network
and web security vulnerabilities. Work in this category mainly includes 
proposals
to improve network intrusion detection~\cite{intrusion_1,intrusion_3},
malware classification~\cite{malware_1, malware_5, malware_6}, 
and password-based
authentication systems~\cite{pw_guessing}.

Similar to our work, a few studies also investigate the feasibility of attacking
online systems using Deep Neural Networks. This includes proposals to
automatically solve CAPTCHAs using Convolutional Neural
Networks~\cite{attack_CAPTCHA, attack_CAPTCHA_2}, and automatically generate
malware domains using Generative Adversarial Networks~\cite{gen_domain}. To the
best of our knowledge, our work is the first to explore attacks on online review
systems using Deep Neural Networks.

\subsubsectitle{Crowdturfing and Review Spam Detection.} Prior work %
characterized crowdturfing marketplaces that supply human labor to enable
attacks on a variety of online platforms, such as review systems, social media,
and search engines~\cite{crowdturfing_1,crowdturfing_4, crowdturfing_5}. 
In addition, work by Wang~\etal investigates
detection of malicious crowdworkers using machine learning and explores the
robustness of machine learning classifiers against evasive tactics by
crowdworkers. Furthermore, researchers have extensively studied detection of 
opinion
spam or fake reviews in online review systems using features based on review
content~\cite{spam_1, spam_3, mukherjee2013yelp} 
and a variety of
metadata~\cite{spam_ling_1, fei2013exploiting, spam_2}.
We differ from these studies by focusing on automatically generating fake
reviews that can evade detection by advanced machine learning classifiers and
human investigation. %

%% file: discussion.tex
\section{Discussion \& Conclusion}
\label{sec:discussion}

In this work, we focus on the potential for misuse of deep learning models in
the context of attacking online review platforms. Our work shows how RNNs can
generate deceptive yet realistic looking reviews targeting restaurants on
Yelp. An extensive evaluation of the quality of generated reviews indicates
the difficulty in detecting such reviews using existing algorithmic
approaches, and even by human examination (which serves as an end-to-end test
of our attack).

We propose a novel approach to defend against RNN-based fake reviews, by
leveraging a fundamental limitation of an RNN-based model: \textit{information
loss incurred during the training process when fitting a large training dataset
to a fixed size statistical model.} Due to the information loss, generated
reviews diverge from real reviews when comparing their character-level
distribution, even when higher level linguistic characteristics are preserved. Our
scheme, based on supervised learning, can detect machine-generated reviews with
high accuracy (F-score ranging from 0.8 to 0.98 depending on the amount of
available ground-truth) and outperforms existing ML-based fake review filters.

\subsubsectitle{Future Work.}
In terms of potential future work, one direction is to consider the role that user and 
content metadata can play in both the attack and defense perspectives. Metadata can be crucial in 
terms of deceiving users (\eg by increasing the number of friends/contacts on the site) and in 
assisting defenses~\cite{bimal_cosn, spam_ling_1, fei2013exploiting, spam_2,spam_5, spam_4, spam_9} (\eg by 
analyzing the patterns in timestamps of user activites). Orchestrating the general behavior of user 
accounts using deep learning to bypass metadata based defenses could be an interesting research 
challenge.
Second, while we limit ourselves to the domain of online review systems and fake review
attacks, deep learning-based generative text models can be applied to launch attacks in
other scenarios as well. We highlight two of these possible application scenarios.

\vspace{0.1in}
\noindent \emph{Strengthening Sybil Attacks.} Attackers can use our
techniques to generate realistic looking text-based user behavior
patterns~\cite{ai_attack1}, \eg posting, commenting and messaging. This can
help attackers make Sybil (fake) accounts indistinguishable
from legitimate accounts based on
textual content. A special case of this involves launching an impersonation
attack in online social networks~\cite{impersonate_imc}.

\vspace{0.1in}
\noindent \emph{Fake News Generation.} Identifying fake news, \ie ``a made-up
story with an intention to deceive''~\cite{fake_news_media}, currently
remains an open challenge~\cite{fake_news}. The research community has
started to explore the possibility of automating the detection process by
building an AI-assisted fact-checking pipeline~\cite{fakenewscheck1,
  fakenewscheck2, williamfakedata}. We believe that AI can not only assist
fake news detection but also generate fake news. Given the availability of
large-scale news datasets~\cite{news_dataset}, an attacker can potentially
generate realistic looking news articles using a deep-learning approach
(RNN). And due to its low economic cost, the attacker can pollute social
media newsfeeds with a large number of fake articles.

We hope our results will bring more attention to the problem of malicious
attacks based on deep learning language models, particularly in the context
of fake content on online services, and encourage the exploration and
development of new defenses.

%% file: appendix.tex
\section{Review Customization Details}
\label{appendix:context}
We show the details of review customization process 
(Section~\ref{subsec:method}) in 
Algorithm~\ref{alg:context}.

\begin{algorithm}[!b]
	\begin{algorithmic}[1]
		\LeftComment{input: R-initial review, T-reference review set, 
			C-topic keyword, $\text{MIN}_\text{sim}$-similarity threshold}
		\Procedure{Review Customization}{R, T, C, $\text{MIN}_\text{sim}$}
		\LeftComment{find nouns in R close to C}
		\State P $\leftarrow$ $\emptyset$
		\For{t $\in$ T}
		\For{n $\in$ Nouns(t)}
		\If{similarity(n, C) $>$ $\text{MIN}_\text{sim}$}
		\State P$\leftarrow$ P  $\bigcup$  n
		\EndIf
		\EndFor
		\EndFor
		\LeftComment{find nouns in R also close to C, replace by sampling 
		from P}
		\For{n $\in$ Nouns(R)}
		\If{similarity(n, C) > $\text{MIN}_\text{sim}$}
		\State S $\leftarrow$ $\emptyset$
		\For{p $\in$ P}
		\State S $\leftarrow$ S $\bigcup$ similarity(n, p)
		\EndFor
		\State z $\leftarrow$ softmax(S)
		\State $\text{n}^*$ $\leftarrow$ sample from P based on z
		\State replace n by $\text{n}^*$
		\EndIf
		\EndFor
		\State return R
		\EndProcedure
	\end{algorithmic}
	\caption{Review Customization}
	\label{alg:context}
\end{algorithm}

\section{More Examples of the Generated Fake Reviews}
\label{appendix:mg_example}
We present examples of generated reviews targeting a specific restaurant
(description shown below). The samples are generated under different temperatures 
and from the models trained using five-star, three-star and one-star Yelp 
training review corpus.\\

\SetBlockThreshold{0} 
\begin{small}
Target Restaurant Description
\begin{scriptsize}
\blockquote{Experience MKT, a market-to-table San Francisco restaurant and bar. 
Taste your way through ultra-creative handcrafted cocktails, small plates for 
sharing, and steaks from the finest California ranches. Meet friends and relax 
in our retro urban-chic atmosphere. In the heart of SoMa - with panoramic 
fifth-floor views of Market Street and downtown - this is one of San 
Francisco's best restaurant venues for social drinks and dining.}
\end{scriptsize}

\subsection{Machine-generated One-Star Reviews}

\noindent Generated with \textit{temperature} 0.1
\begin{scriptsize}

\blockquote{
	I was so excited to try this place out for the first time and the food was 
	awful. I ordered the chicken sandwich and it was so salty that I could not 
	eat it. I was so disappointed that I was so disappointed in the food. I was 
	so disappointed that I was so disappointed with the service.
}
\end{scriptsize}

\noindent Generated with \textit{temperature} 0.3
\begin{scriptsize}
\blockquote{
I had a drink at this restaurant. I was there for a couple of months. The 
restaurant was clean but the food was not good. The service was very slow and 
the service is so bad. I would not recommend this place to anyone.
}
\end{scriptsize}

\noindent Generated with \textit{temperature} 0.5
\begin{scriptsize}
\blockquote{
The worst place I've ever been too. I waited for a while and the place was not 
available and the service was absolutely terrible. The server was rude and not 
apologetic and unprofessional. I will not be going back to 
this location.}
\end{scriptsize}

\noindent Generated with \textit{temperature} 0.7
\begin{scriptsize}
\blockquote{
I have been here twice, very expensive for what you get. We stayed here before 
at our table and were 
seated in a line for 15 minutes and the restaurant was full and the cashier 
recommended the drinks and they forgot the cookie cover charge and we were told 
it was not an issue. I would never recommend this place to anyone.
}
\end{scriptsize}

\noindent Generated with \textit{temperature} 1.0
\begin{scriptsize}
\blockquote{DO NOT WASTE YOUR TIME AND MONEY! The absolute worst service I have 
ever experienced. This place is a joke. The waitress was rude and said she 
would put the manager to come out but never happened. I wish I could give zero 
star.}
\end{scriptsize}

\subsection{Machine-generated Three-Star Reviews}

\noindent Generated with \textit{temperature} 0.1
\begin{scriptsize}
\blockquote{
The food was good but not great. I had the lasagne and the sauce was a 
little too sweet for my taste.  The salad was good but the sauce was a little 
too sweet for my taste. The sauce was a little too salty for my tastes.  The 
sauce was a little too sweet for my taste.
}
\end{scriptsize}

\noindent Generated with \textit{temperature} 0.3
\begin{scriptsize}
\blockquote{
I've had the fried chicken and the chicken breast 
with chicken soup with sweet potato fries.  I don't know what a lot of the meat 
was on the salty side as well. The sauce was a bit salty and didn't have much 
flavor.  The salad was tasty and well presented but the bread was super tasty.
}
\end{scriptsize}

\noindent Generated with \textit{temperature} 0.5
\begin{scriptsize}
\blockquote{
I was here for a weekend brunch and the food was ok. I love the pizza 
that is a chain restaurant. I think the service is excellent. I had the 
spaghetti and they were very good and the hot dog was good. I got the 
red velvet chocolate cake special which was very good but the service was a 
little slow. The food was good, but not up to par with other places nearby.
}
\end{scriptsize}

\noindent Generated with \textit{temperature} 0.7
\begin{scriptsize}
\blockquote{
The food wasn't bad.  The cupcakes are okay and the service is excellent but 
the prices are a bit high.  I do like the fresh made salad and drink 
specials. I would recommend this place for a place to grab a 
bite for a couple of times.
}
\end{scriptsize}

\noindent Generated with \textit{temperature} 1.0
\begin{scriptsize}
\blockquote{
Came here for lunch today and the place was pretty empty. The steak was good 
but the chicken they had a little less oily and overcooked. I would recommend 
this place if you are looking for a cheap place to stop by.
}
\end{scriptsize}

\subsection{Machine-generated Five-Star Reviews}

\noindent Generated with \textit{temperature} 0.1
\begin{scriptsize}
\blockquote{
I have been going to this place for a few years now and I have never had a bad 
experience. The service is great! They are always so friendly and helpful. I 
will definitely be back and I will be back for sure!
}
\end{scriptsize}

\noindent Generated with \textit{temperature} 0.3
\begin{scriptsize}
\blockquote{
This place is amazing! The bartenders are absolutely amazing. The pasta is 
delicious and I love their pastries and it is amazing. I love the breakfast, 
friendly staff and the price is very reasonable. I have never had a bad 
experience here. I will be back for sure!
}
\end{scriptsize}

\noindent Generated with \textit{temperature} 0.5
\begin{scriptsize}
\blockquote{
I love this place. I went with my brother and we had the vegetarian pasta and 
it was delicious. The beer was good and the service was 
amazing. I would definitely recommend this place to anyone looking for a great 
place to go for a great breakfast and a small spot with a great deal.
}
\end{scriptsize}

\noindent Generated with \textit{temperature} 0.7
\begin{scriptsize}
\blockquote{
I have been a customer for about a year and a half and I have nothing but great 
things to say about this place.  I always get the pizza, but the 
Italian beef was also good and I was impressed. The service was outstanding. 
The best service I have ever had. Highly recommended.
}
\end{scriptsize}

\noindent Generated with \textit{temperature} 1.0
\begin{scriptsize}
\blockquote{
The food here is freaking amazing, the portions are giant. The cheese 
bagel was cooked to perfection and well prepared, fresh \& delicious! The 
service was fast. Our favorite spot for sure! We will be back!
}
\end{scriptsize}

\end{small}

\begin{figure}[!t]
	\centering 	    		 
	\subfigure[Examining human performance on machine-generated review 
	detection.]
	{\includegraphics[width=0.445\textwidth]{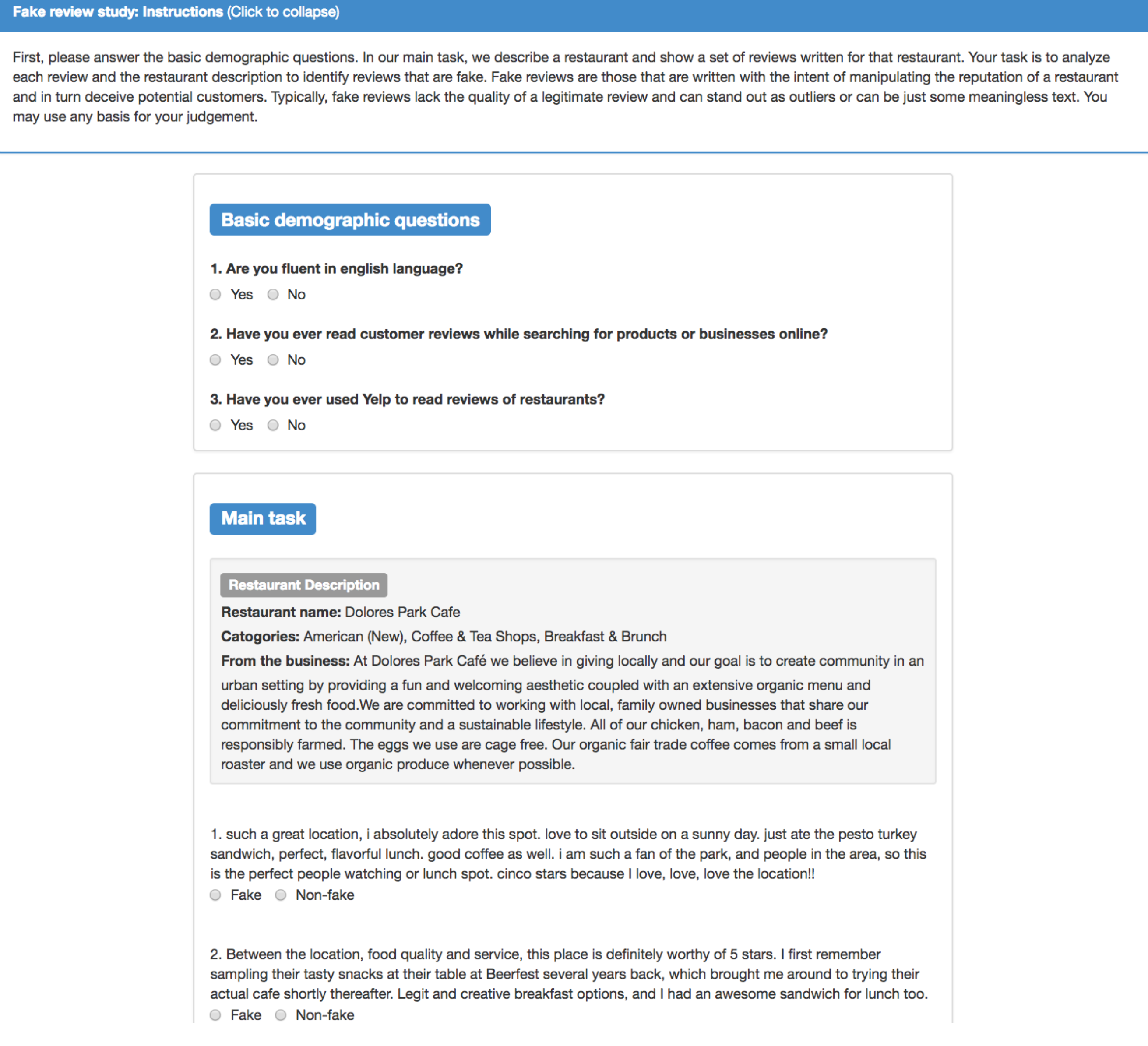}}
	\hfill     		
	\subfigure[Collecting helpfulness rating of the machine-generated reviews.]
	{\includegraphics[width=0.445\textwidth]{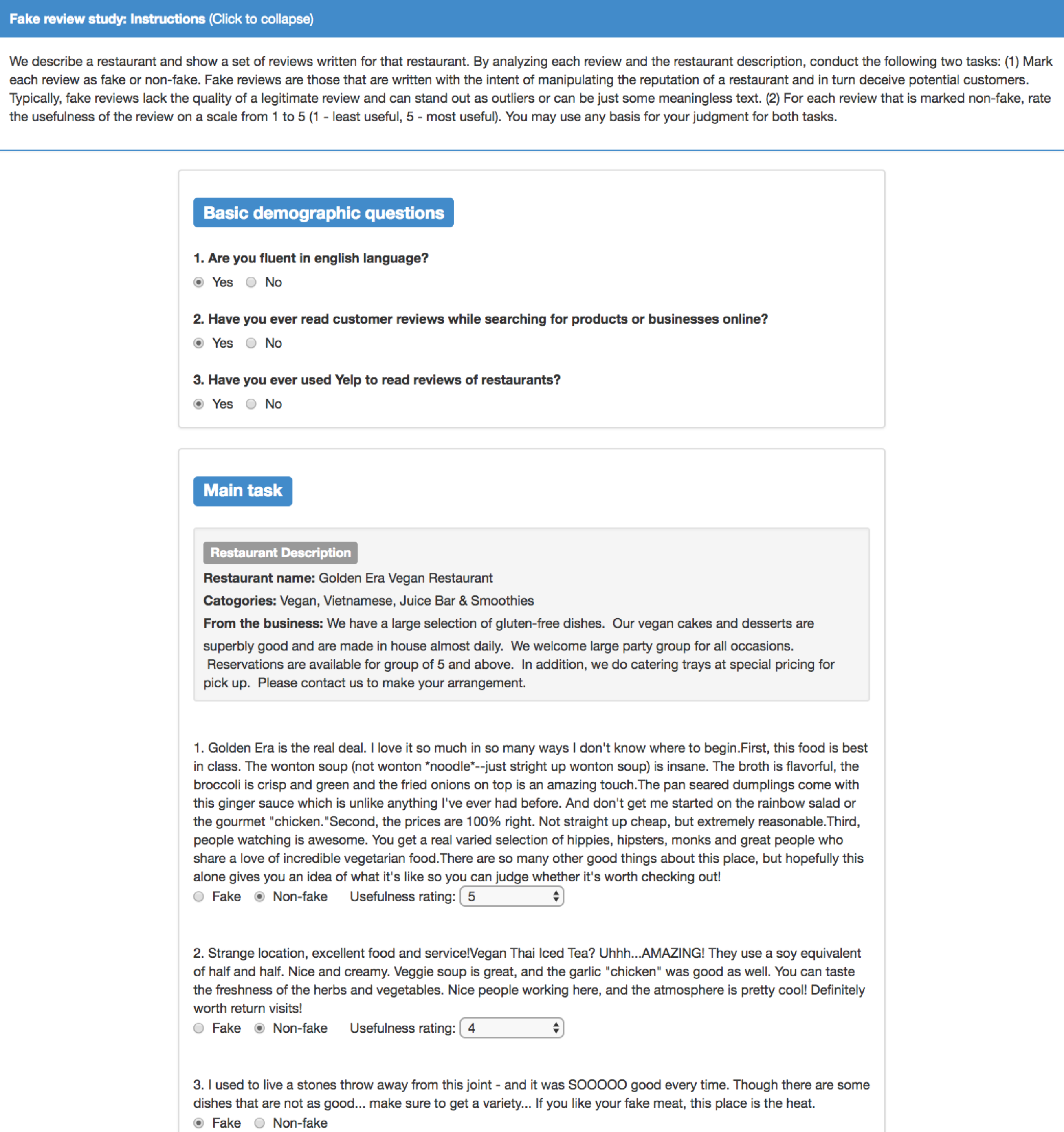}}
	 \caption{Screenshots of the survey designed for end-to-end user study.}
	\label{fig:suverys}
\end{figure}

\section{Screenshots of User Study Surveys}
\label{appendix:survery}
\subsection{Fake/Real Review Detection}
Figure~\ref{fig:suverys}(a) shows a screenshot of the survey designed for 
examining human performance on machine-generated review detection.

\subsection{Review Helpfulness Rating}
Figure~\ref{fig:suverys}(b) shows another screenshot of the second round survey 
designed for collecting the helpfulness score of the machine-generated reviews.